\documentclass[aps,pra,twocolumn,groupedaddress,longbibliography]{revtex4-1}
\usepackage[dvipsnames]{xcolor}
\definecolor{mblue}{RGB}{31, 119, 180}
\usepackage{hyperref}
\hypersetup{backref,pdfpagemode=FullScreen,colorlinks=true,breaklinks,urlcolor=mblue,linkcolor=mblue,citecolor=mblue}
\usepackage{physics, braket, bm, amsthm, amsmath, amssymb, mathrsfs, graphicx, dcolumn}

\begin{document}
	
	\title{Detection of entangled states supported by reinforcement learning}
	
	

	\author{Jia-Hao Cao}
	\thanks{These authors contributed equally to this work.}
	\affiliation{State Key Laboratory of Low Dimensional Quantum Physics, Department of Physics, Tsinghua University, Beijing 100084, China}
	
	\author{Feng Chen}
	\thanks{These authors contributed equally to this work.}
	\affiliation{State Key Laboratory of Low Dimensional Quantum Physics, Department of Physics, Tsinghua University, Beijing 100084, China}
	
	\author{Qi Liu}
	\thanks{Present address: Laboratoire Kastler Brossel, Coll\`{e}ge de France, CNRS, ENS-PSL University, Sorbonne Universit\'{e}, Paris, France.}
	\affiliation{State Key Laboratory of Low Dimensional Quantum Physics, Department of Physics, Tsinghua University, Beijing 100084, China}
	
	\author{Tian-Wei Mao}
	\affiliation{State Key Laboratory of Low Dimensional Quantum Physics, Department of Physics, Tsinghua University, Beijing 100084, China}
	
	\author{Wen-Xin Xu}
	\affiliation{State Key Laboratory of Low Dimensional Quantum Physics, Department of Physics, Tsinghua University, Beijing 100084, China}
	
	\author{Ling-Na Wu}
	\email{lingna.wu@hainanu.edu.cn}
	\affiliation{Center for Theoretical Physics and School of Science, Hainan University, Haikou 570228, China}
	
	\author{Li You}
	\email{lyou@tsinghua.edu.cn}
	\affiliation{State Key Laboratory of Low Dimensional Quantum Physics, Department of Physics, Tsinghua University, Beijing 100084, China}
	\affiliation{Frontier Science Center for Quantum Information, Beijing, China}
	\affiliation{Beijing Academy of Quantum Information Sciences, Beijing 100193, China}

	\begin{abstract}
		Discrimination of entangled states is an important element of quantum enhanced metrology. This typically requires low-noise detection technology. Such a challenge can be circumvented by introducing nonlinear readout process. Traditionally, this is realized by  reversing the very dynamics that generates the entangled state, which requires 
		a full control over the system evolution. In this work, we present nonlinear readout of highly entangled states by employing reinforcement learning~(RL) to manipulate the spin-mixing dynamics in a spin-1 atomic condensate. The RL found results in driving the system towards an unstable fixed point, whereby the (to be sensed) phase perturbation is amplified by the subsequent spin-mixing dynamics.
		Working with a condensate of 10900 $^{87}$Rb atoms, we achieve a metrological gain of $6.97^{+1.30}_{-1.38}$ dB beyond the classical precision limit. 
		Our work would open up new possibilities in unlocking the full potential of entanglement caused quantum enhancement  in experiments.

	\end{abstract}
	
	\maketitle
	
\begin{figure} [htp!]
		\includegraphics[width=\columnwidth]{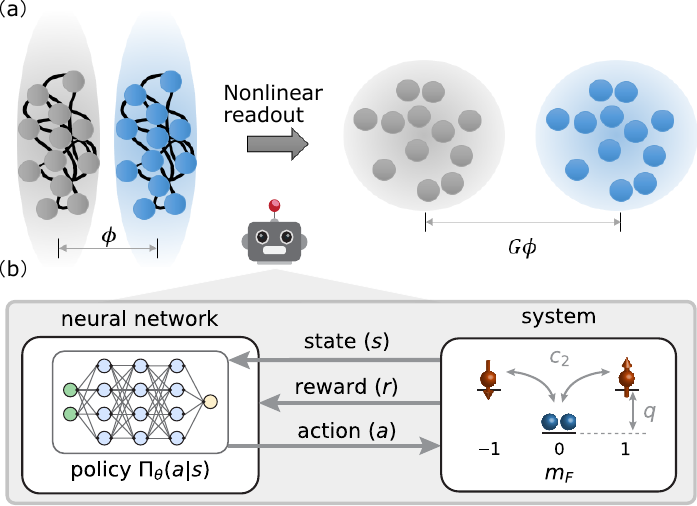}
		\caption{
			(a) The precision of measurements is constrained by the quantum fluctuations of probe states, as depicted by the shaded region in the figure. Entangled states with squeezed fluctuation distributions~(the fluctuation along the phase-encoding direction is squeezed, as a cost, the fluctuation along the orthogonal direction is enlarged) allow for more precisely estimating parameter $\phi$ than classical resources (left panel).
			The benefit of squeezed noise, however, can be easily overshadowed by detection noise, which, combined with quantum fluctuations, establishes the overall noise level.  
			By introducing a nonlinear readout process~(right panel), which preferentially amplifies signal over noise, noise-robust detection with high phase sensitivity can be realized· This work exploits RL to guide nonlinear readout of highly entangled states.
			(b) A typical interaction loop between neural network and  system for RL training.
			In this work, the system is a spin-1 condensate. The QZS $q$ is modulated according to the policy from RL to manipulate the dynamics.
		}
		\label{fig1}
	\end{figure}

	\textit{Introduction.---}
	Entanglement plays a crucial role in quantum metrology~\cite{giovannetti2004,luca2018}, which aims at beating the standard quantum limit (SQL) or classical limit achievable with uncorrelated particles by using quantum resources.
	To fully harness the quantum advantages offered by entangled states, it is necessary to discriminate between entangled states with and without parameter perturbations. However, detecting entangled states is often vulnerable to technical noise, compromising the metrological advantage derived from using entanglement. To overcome this challenge, nonlinear readout techniques have been introduced~\cite{emily2016,frowis2016,macri2016}, based on amplifying the distinction between perturbed and unperturbed entangled states by disentangling the particles [see Fig.~\ref{fig1}(a)]. Such a nonlinear process exhibits high sensitivity to perturbations, producing markedly different output states even with minute perturbations, and thus promises quantum-enhanced sensing via signal amplification.

	A straightforward implementation of nonlinear readout of entangled states comes from time-reversing the very dynamics that generates the entanglement in the first place~\cite{hosten2016quantum,burd2019,gilmore2021,colombo2021,linnemann2016,hudelist2014,Li2018}. Such a procedure, however, requires a precise knowledge of the history of the dynamical process, as well as a full control over the system, which is 
	highly nontrivial for a general many-body quantum
	system. 
	In this work, we demonstrate an easily implementable nonlinear readout process supported by reinforcement learning~(RL)~\cite{sutton2018} that requires no knowledge of entanglement generation history and is realized by modulating only a linear control field. 
	
	As an important branch of machine learning,
	RL targets an optimal strategy for accomplishing a specific task without prior knowledge. 
	It optimizes decision making based on interacting with the system instead of hints or guidances to the solutions. 
	In quantum science and technology, RL has been applied to quantum state engineering~\cite{bukov2018,dalgaard2020,wauters2020,guo2021,yao2020a,Haug_2020,borah2021,porotti2022deep}, quantum control~\cite{an2019,niu2019,wang2020,PhysRevA.101.052327,Saggio2021}, quantum metrology~\cite{Xu2019,Schuff_2020,PhysRevResearch.3.033279,2022arXiv220300189Q}, quantum error correction~\cite{fosel2018,nautrup2019,Andreasson2019quantumerror}, and quantum compiling~\cite{PhysRevLett.125.170501,Moro2021,He_2021} etc., with great successes. Here we employ RL to guide nonlinear readout of highly entangled states for quantum enhanced sensing.

\begin{figure*}
	\centering
	\includegraphics[width=1.99\columnwidth]{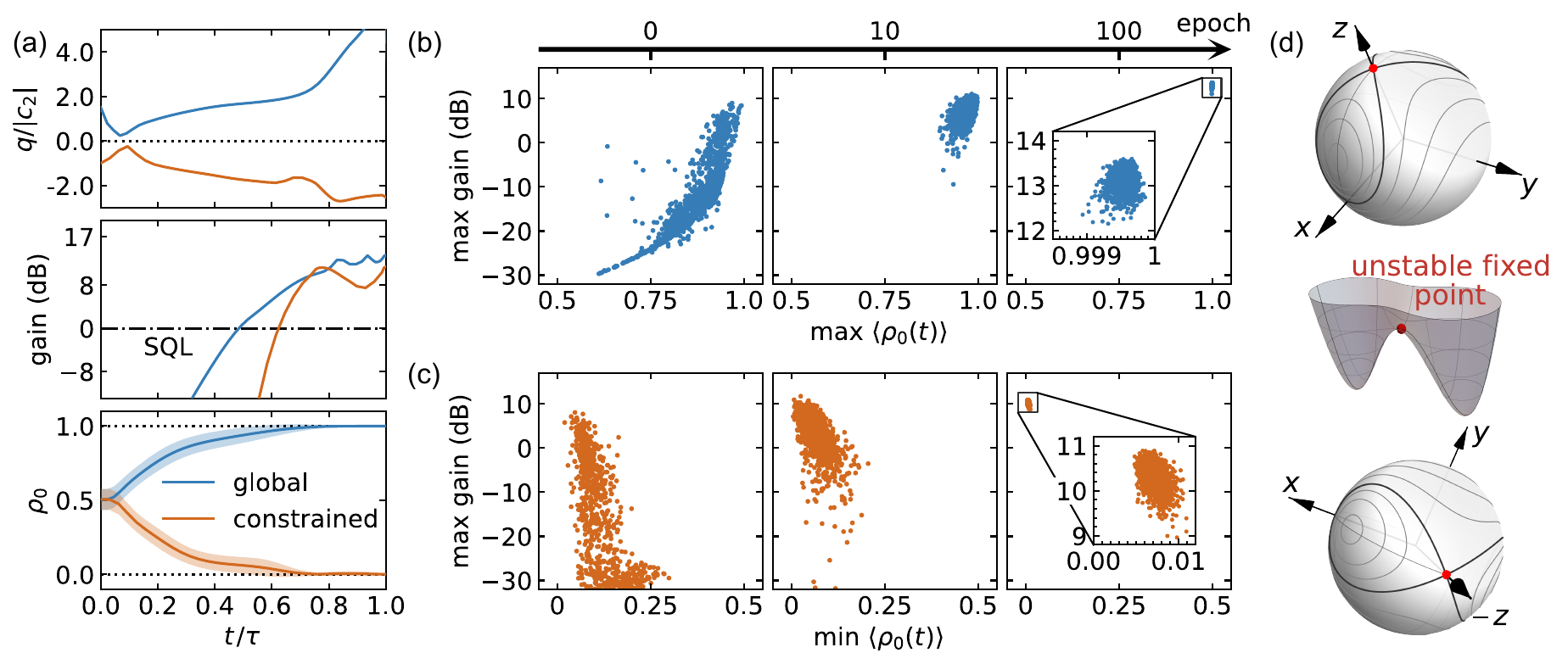}
	\caption{{(a) RL training results for the readout operation in a small system of $N=50$ particles, with the $q$ profile, the metrological gain, and fractional population $\rho_0$ evolution~(in the absence of phase encoding) shown from top to bottom. The blue~(orange) data denotes the results for a global~(constrained) search without~(with) constrains~($q<0$) on $q$. The shaded regions denote the fluctuation of $\rho_0$, $\Delta \rho_0$. 
		(b) Maximal achievable metrological gain vs maximal accessible $\rho_0$ during the readout process from $2000$ trajectories for each sampled according to the policy at $0$, $10$ and $100$ epoch. As the learning proceeds~(from left to right), the readout operation tends to drive the system towards the polar state with a large $\langle \rho_0\rangle$ for getting a large metrological gain.
		(c) Similar results as (b), but for the case with $q<0$ constrain. 
		(d) Spin-nematic sphere~\cite{PhysRevLett.111.090403} with energy contours for $q=|c_2|$~(top) and $q=-|c_2|$~(bottom). The three axes are $x=\sqrt{1-z^2}\cos(\theta/2)$, $y=\sqrt{1-z^2}\sin(\theta/2)$, and $z=2\rho_0-1$, with $\theta$ being the spinor phase. Middle panel: energy surface. The polar state~(marked with a red sphere) sits at the north pole, which is a saddle point for $0<q<2|c_2|$. The south pole becomes an unstable fixed point when $-2|c_2|<q<0$.}
	}
	\label{fig2}
\end{figure*}

	Our system is composed of a $^{87}$Rb atomic condensate in the ground hyperfine $F=1$ manifold $(m_F=0,\pm1)$, described by the Hamiltonian ($\hbar=1$ hereafter),
	{\small
		\begin{eqnarray}
		\label{eq_ham}
		H &=& \frac{c_2}{2N}{\left[\left(2a_0^\dag a_0^\dag a_1 a_{-1}+\rm{h.c.} \right)+(2N_0-1)(N-N_0)\right]}  -  q(t)N_0, \notag
	\end{eqnarray}}
under the assumption of the same spatial mode for the three spin components~\cite{law1998}.
	Here $a_{m_F}$~($a_{m_F}^\dag$) denotes annihilation (creation) operator and 
	$N_{m_F}=a^{\dagger}_{m_F}a_{m_F}$ counts the number of atoms in $m_F$ component with $N = \sum_{m_F} N_{m_F}$ the total  number of atoms. The Hamiltonian can also be written as $H = \frac{c_2}{2N}{\bf L}^2 - q(t)N_0$ in terms of collective spin ${\bf L} \equiv \sum_{\mu, \nu}  a^{\dagger}_{\mu}{\bf F}_{\mu\nu}a_{\nu}$ with ${\bf F}_{\mu\nu}~(\mu,\nu=-1,0,+1)$ the spin-1 matrix element. It conserves magnetization $L_z = N_{+1} - N_{-1}$, as well as the total particle number $N$.
	The first term in the Hamiltonian describes spin-exchange interaction of ferromagnetic type at strength $c_2 < 0$, whereby atoms in $m_F=0$ are transferred to $m_F = \pm 1$ in pairs and vice versa.
	The second term describes an effective quadratic Zeeman shift~(QZS) $q$ which can be tuned experimentally~\cite{gerbier2006}. The competition between the above interactions gives rise to intriguing spin-mixing dynamics~\cite{law1998}. In this work, $q$ is modulated according to the policy from RL to manipulate the dynamics.

	\begin{figure*}[htp!]
		\centering
		\includegraphics[width=2\columnwidth]{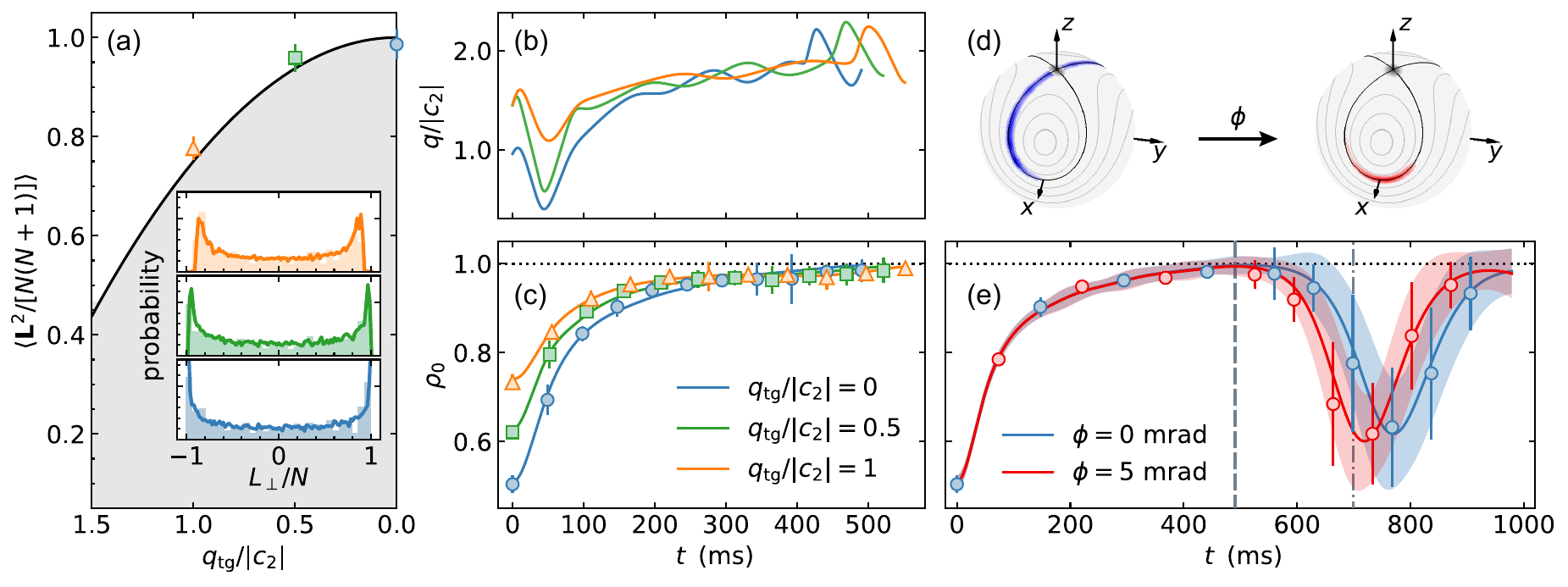}
		\caption{ 
			(a) The collective spin length of the spinor BEC. The black solid line (gray shaded region) denotes theoretical results of the ground state~(excited states). The markers denote results of the correspondingly prepared target probe states at $q_{\rm tg}/|c_2| = 0$ (blue), $0.5$ (green), and $1$ (orange) from 500 continuous experimental runs on measuring transverse spins~\cite{supp}, whose distributions are shown in the insets. 
			The solid lines in the insets come from numerical simulations.
			(b) The RL-learnt $q$ profile for the nonlinear readout of the three probe states. (c) The evolution of $\rho_0$ during the readout process.  (d) Illustration of states at the end of the RL-learnt readout operation~[$t=491$~ms, marked by vertical dashed line in (e)]~(gray data) and at the detection time~[$t=699$~ms, marked by vertical dot-dashed line in (e)]~(colored data) on the spin-nematic sphere. The left~(right) panel denotes the case when no~(a) phase is encoded into the probe state. (e) The evolution of $\rho_0$ when no phase~(blue data) or a phase~(red data) is encoded to the balanced spin-1 Dicke state prepared at $t=0$. The QZS is $q=|c_2|$. For (c) and (e), markers are data from
			$100$ experimental runs. The solid lines (shaded regions)
			denote numerically computed mean (uncertainty).
		}
		\label{fig3}
	\end{figure*}
	\textit{Reinforcement learning.---}
	In a typical RL task~[see Fig.~\ref{fig1}(b)], the agent learns from trial and error to achieve a pre-specified goal. By inspecting the system through some observables (state), the agent makes decisions (action) according to certain tactic (policy, a mapping between state and action) to alter the state of the system. 
	At the same time, it collects a feedback (reward) from the system, which measures whether the decision is constructive or not. 
	After many rounds of such interactions, the agent refines the tactic based on the collected information and updates continuously until it gains sufficient experience to arrive at an optimal tactic for achieving the goal~(see supplemental material~\cite{supp} for more details).

	We first benchmark the RL training in a small system with $N = 50$ atoms. In the quantum metrology application concerned here, our aim is to precisely detect a phase $\phi$ which is encoded by a rotation operation described by $U_\phi=e^{-i\phi L_x}$ with $L_x=(a_1^\dag a_0 + a_0^\dag a_{-1} + \rm{h.c.})/\sqrt{2}$.
	Starting from the polar state with all atoms in $m_F=0$ component, the probe state is prepared by targeting the balanced spin-1 Dicke state~\cite{guo2021,supp}, which is the ground state of the system at $q=0$. Instead of direct detection after phase encoding~\cite{zou2018}, 
	we employ RL to guide a nonlinear readout operation before the measurement, thereby inducing significant distinctions between the final states evolved from without and with encoding rotation $U_\phi$.  The encoded phase is extracted by measuring the fractional population in $m_F=0$ component, $\rho_0$. Inferred from error propagation, the phase sensitivity of $\phi$ reads
	\begin{eqnarray}
		\label{ps}
		\Delta\phi = \Delta \rho_0/\abs{\partial_{\phi}\langle \rho_0\rangle},
	\end{eqnarray}
	which is determined by the fluctuation of $\rho_0$ and the slope of its mean with respect to $\phi$. The reward of RL training is set to maximize the metrological gain, $-20\log_{10}(\Delta\phi/\Delta \phi_{\rm SQL})$, over the three-mode SQL $\Delta \phi_{\rm SQL}=1/(2\sqrt{N})$.
	We present the results of RL training for a global search~(without constrains on $q$) in Fig.~\ref{fig2}(a) by blue data, with the $q$ profile for the nonlinear readout process, the corresponding metrological gain, and  $\rho_0$ evolution~(in the absence of phase encoding) shown from top to bottom. From the $\rho_0$ evolution, one can see that the learnt readout operation drives the system towards the initial polar state where all atoms occupy the $m_F=0$ component~($\rho_0=1$). 
Similar behavior is found when noises due to experimental imperfections are taken into account~\cite{supp}. 
	
	The pivotal role played by the polar state in achieving high phase sensitivity is further elaborated in the learning progress of readout operation, see Fig.~\ref{fig2}(b). Here, we inspect the metrological performance of the readout processes governed by a collection of $q(t)$ profiles.
	For each $q(t)$ profile, one can extract the maximal $\rho_0$ during the readout process and the maximal metrological gain. A summary of them from 2000 trajectories each for three different training epochs~(0, 10, 100) is shown in Fig.~\ref{fig2}(b). 
	As the learning proceeds, the distribution shrinks towards the upper right corner where achieving enhanced metrological performance and passing through the initial polar state~(large maximal $\rho_0$) are found to be strongly correlated. 

	One can understand the observed behavior by inspecting the system from  phase space spanned by mean fields~\cite{PhysRevLett.111.090403}, $\rho_0$ and the spinor phase $\theta=\theta_1+\theta_{-1}-2\theta_0$, where $\theta_{m_F}$ denotes the phase of the $m_F$ component. These two variables are constrained by the single particle mean field energy $\varepsilon=c_2\rho_0(1-\rho_0)(1+\cos\theta)-q\rho_0$. The phase space for $|q|<2|c_2|$~[see top panel of Fig.~\ref{fig2}(d)] is divided by a separatrix~(thick black line). The polar state sits at the north pole, which is an unstable fixed point of the system for $0<q<2|c_2|$~[see the energy surface in the middle panel of Fig.~\ref{fig2}(d)], rendering the 
	 dynamics starting from it highly susceptible to perturbations. This explains why the readout operation drives the system towards the dynamically unstable polar state.	
	
	To justify our interpretation, we perform another RL training, with $q$ constrained in the negative regime, where the unstable fixed point shifts to the south pole~[see bottom panel of Fig.~\ref{fig2}(d)]. If our above understanding is correct, the optimal strategy is expected to drive the system towards the south pole with $\rho_0=0$. This is indeed what we observe, as shown by the orange data in Fig.~\ref{fig2}(a) and (c). {It is worthy to point out  these results suggest that for realizing nonlinear readout it is in fact not necessary to return to the starting point of entanglement generation, which was unanimously believed as a guiding principle in traditional protocols~\cite{emily2016,frowis2016,macri2016,hosten2016quantum,burd2019,gilmore2021,colombo2021,linnemann2016,Li2018,liu2021}. Such an interesting finding offers an alternative to implementing nonlinear readout.
	} 
	
	We make use of the above observation to simplify the training task for larger systems. 
	The calculations of phase sensitivity~\eqref{ps} involve time-consuming computations of various output states under a span of encoded phases.
	To simplify the numerics, we replace the training task for large systems by targeting the initial polar state. 
	To further speed up training and avoid sparse reward, we adopt transfer learning~\cite{supp}, with experience gained from trained neural networks for smaller sized systems in the absence of atom loss applied as initial values to larger or/and realistic experimental systems with loss. This benefits from the remarkable generalization ability of the RL policy~\cite{guo2021}.
	For treating large systems including atom loss, truncated Wigner approximation method~\cite{liu2021, supp, steel1998, sinatra2002, norrie2006, opanchuk2012, drummond2017, johnson2017, gerving2012, hamley2012} is used in our numerical simulations.

	\textit{Experimental implementation.---}
	The RL protocols are implemented experimentally in a spinor condensate of about $10900$ $^{87}$Rb atoms at $c_2 = -2\pi\times2.6(2)$ Hz. 
	The net QZS $q=q_{\rm B} + q_{\rm MW}$ includes a magnetic field contribution $q_{\rm B}$, and a microwave dressing field (detuned from the $|F=1, m_F=0\rangle$ to $|F=2, m_F=0\rangle$ clock transition) contribution $q_{\rm MW}$~\cite{zhao2014}. 
	The bias magnetic field is set at $0.537$~G corresponding to $q_{\rm B} \sim 8 |c_2|$ as a tradeoff between minimizing the influence of radio-frequency (RF) noise and maintaining the stability of QZS~\cite{supp}.
	The experiment starts from a polar state BEC at $q \sim 17 |c_2|$, followed by quenching $q$ to $1.5|c_2|$ through tuning the microwave power before the state generation operation. Subsequently, $q$ is ramped to prepare the probe states~\cite{supp}, which serve as the input states for the nonlinear readout operations. Here three different probe states are studied, prepared by targeting the ground states of the system at $q_{\rm tg}=0$, $0.5|c_2|$ and $|c_2|$, respectively. 
	The metrological potential of a probe state $|\psi_p\rangle$ is limited by the quantum Cram\'er-Rao bound~\cite{luca2018} to, $(\Delta \phi)^2 \geqslant 1/F_Q(\ket{\psi_p}, L_x)$, with the quantum Fisher information~(QFI) $F_Q(\ket{\psi_p}, L_x) =4(\Delta L_x)^2= 4\langle L_x^2 \rangle = 2\langle {\bf L}^2\rangle$ in the $L_z=0$ subspace for a pure state. 
	In our system, the ground state possesses larger collective spin length $\langle {\bf L}^2\rangle$ than excited states, and $\langle {\bf L}^2\rangle$ increases when $|q|$ decreases~[see Fig.~\ref{fig3}(a)], with the balanced spin-1 Dicke state at $q=0$ offering the highest QFI, and thus the highest theoretical phase sensitivity. 
	On the other hand, preparing the ground state by ramping $q$ from $q \gg |c_2|$ to smaller $q_{\rm tg}$ requires longer time~\cite{supp}, and thus suffers more from atom loss. 
	Therefore, it is nontrivial to predict which state provides the highest phase sensitivity in experiment.
	
	Phase encoding is enacted by a well-calibrated Rabi rotation of the state using RF field that SU(2)-symmetrically couples all three $m_F$ components.
	To detect the phase in a noise-robust way, we perform nonlinear readout operation guided by RL. Namely, $q$ is ramped according to the RL profile, as shown in Fig.~\ref{fig3}(b). 
	The evolution of $\rho_0$ during the readout process, measured by absorption imaging after spatial Stern-Gerlach separation~\cite{supp}, is shown in Fig.~\ref{fig3}(c). 
	The experimental data~(markers) shows an excellent agreement with theory~(solid lines). Approaching $\rho_0=1$ at the end of the readout process signals the success of the RL protocol.
	
	To illustrate the sensitivity of the system to phase perturbation, we compare the dynamics of the system when a finite phase~(red data) or no phase~(blue data) is encoded to the probe state~(balanced spin-1 Dicke state). The results are shown in Fig.~\ref{fig3}(d) and (e). Before the system approaches the polar state~($t<491$~ms), one can hardly distinguish these two cases. While a substantial difference is observed after the system passes through the polar state~($t>491$~ms).
	Such a sensitive dependence of $\rho_0$ on phase perturbation exemplifies the  discrimination of the encoded states with slightly different phases. 

		\begin{figure}
		\centering
		\includegraphics[width=\columnwidth]{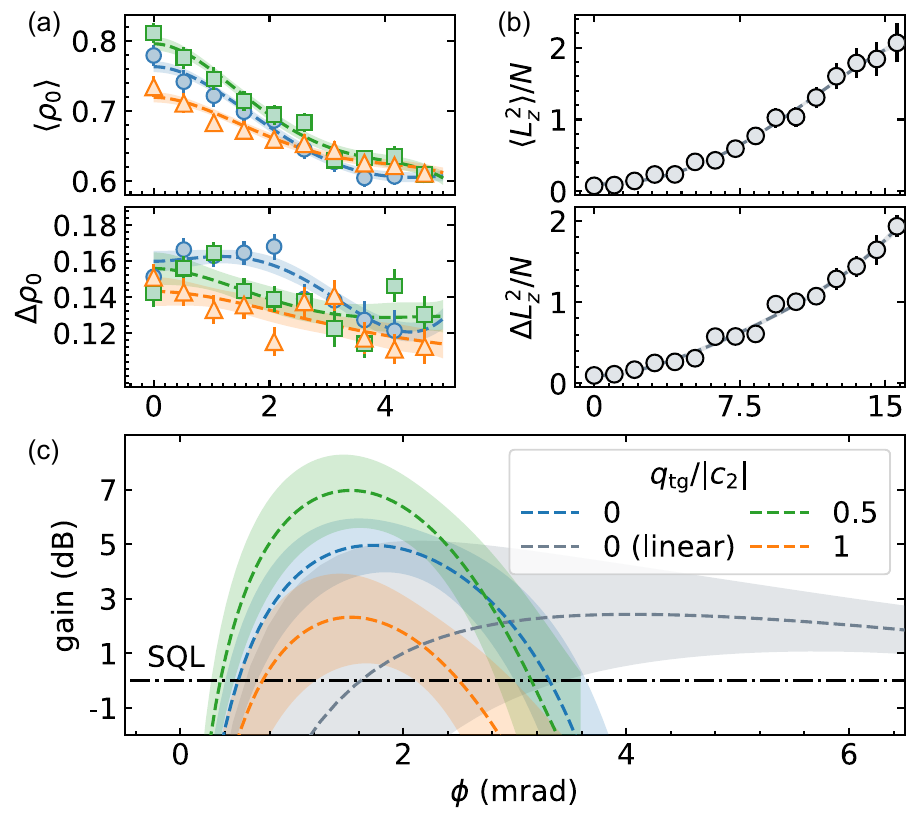}
		\caption{(a) The averages (upper panel) and standard deviations (lower panel) of fractional spin population in $m_F=0$, $\rho_0$, in the vicinity of phase $\phi=0$. Each data point comes from 100 continuous experimental runs. 
			(b) The averages (upper panel) and standard deviations (lower panel) of $L_z^2$, for the balanced spin-1 Dicke state without nonlinear readout operation. Each data point comes from 50 continuous experimental runs. 
			For (a) and (b), dashed lines are fitted results.
			(c) The corresponding metrological gains over three-mode SQL (dot-dashed line).
			The dashed lines are obtained based on error propagation by using the fitted results in (a) and (b). The shaded region indicates the fitting uncertainty.}
		\label{fig_sen}
	\end{figure}

	We show the metrological performance of the three probe states in Fig.~\ref{fig_sen}. The mean value of $\rho_0$ and its fluctuation $\Delta \rho_0$~(measured at $208$~ms after the system approaches the polar state~\footnote{From the training results of small systems where noises are taken into account, we learn that after the system approaches the polar state, an additional period of spin mixing dynamics is needed to submerge the effect of noise by signal amplification. The duration of this extra evolution is optimized in experiments to give the highest metrological gain.}) for small encoded phases are shown in Fig.~\ref{fig_sen}(a), with circles denoting experimental data and dashed lines the fitted curves.  
	Phase sensitivity is calculated based on error propagation~\eqref{ps}
	by using the fitted data~[dashed lines in  Fig.~\ref{fig_sen}(a)] and shown as colored data in Fig.~\ref{fig_sen}(c). The optimal metrological gains for the three states with $q_{\rm tg}/|c_2|=0$, $0.5$ and $1$ are found to be $\xi^2 = -20\log_{10}(\Delta \phi/[1/(2\sqrt{N})]) \simeq 4.96^{+0.96}_{-1.02}$, $6.97^{+1.30}_{-1.38}$ and $2.32^{+1.56}_{-1.73}$ dB, respectively, beyond three-mode SQL~\cite{zou2018} of $1/(2\sqrt{N})$ with the atom number of the probe state $N \simeq 10400$.
	The probe state at $q_{\rm tg}/|c_2| = 0.5$ therefore outperforms due to the aforementioned tradeoff between theoretical metrological gain and atom loss. 	 
	
	For comparison, we also directly detect the phase-encoded spin-1 Dicke state~\cite{supp,zou2018}, i.e., without nonlinear readout process. In this case, the phase can be extracted by measuring $L_z^2$, whose mean values and fluctuations for small phases are shown in Fig.~\ref{fig_sen}(b).
	The inferred phase sensitivity~[gray line in Fig.~\ref{fig_sen}(c)] gives $\xi^2\simeq 2.42^{+1.78}_{-1.84}$ dB beyond three-mode SQL. Hence, an enhancement of $2.5$~dB is observed by introducing the nonlinear readout operation. The main constraint for further improvement on the achievable phase sensitivity is atom loss, which is an extra price to pay for performing the nonlinear readout operation.
	In conclusion, we employ RL to guide nonlinear readout of entangled states based on manipulation of the spin-mixing dynamics of a spin-1 condensate by modulating the QZS. {Remarkably, the optimal strategy is found to drive the system towards an unstable fixed point, where small perturbations grow in time. Our results suggest that there are various approaches to realizing nonlinear readout, besides the paradigmatic one of returning to the starting point of entanglement generation. 
	Our method can be generalized to other systems, see Ref.~\cite{supp} for a discussion of applying our method to a spin-1/2 system. 
	}  
	
	\begin{acknowledgments}
		We thank Dr. M. K. Tey and Dr. J. Yu for helpful discussions.
		This work is supported by the National Natural Science Foundation of China (NSFC)(grants No. U1930201, 11654001, and 91836302), 
		and by the National Key R\&D Program of China (grant No. 2018YFA0306504). 
	\end{acknowledgments}
	
	
%

\pagebreak

\clearpage

\onecolumngrid
\begin{center}
\textbf{\large {Supplemental Material for \\
	``Detection of entangled states supported by reinforcement learning''}}
\end{center}

\setcounter{equation}{0}
\setcounter{figure}{0}
\setcounter{table}{0}
\setcounter{page}{1}
\renewcommand{\theequation}{S\arabic{equation}}
\renewcommand{\thefigure}{S\arabic{figure}}
\renewcommand{\bibnumfmt}[1]{[S#1]}

This supplemental material provides expanded discussions on several issues left out of the main text due to space limitation.
They include: (i) a detailed description of reinforcement learning (RL) task in Sec. \ref{rl} and transfer learning in Sec. \ref{tl};
(ii) simulation of dissipative systems in Sec. \ref{sec2};
(iii) experimental methods in Sec. \ref{sec4}; 
(iv) more data in Sec. \ref{sec5};
(v) a brief introduction to spin-1 Dicke state in Sec. \ref{sec:dicke};
(vi) a comparison of our protocol with time-reversal protocol in Sec. \ref{sec6},
and (vii) a study on nonlinear readout in a spin-1/2 system in Sec.~\ref{spin-half}, which also includes a comparison of RL with traditional optimization methods.

\section{Reinforcement learning}\label{rl}

\subsubsection{Introduction to the training process}

The training process comprises multiple epochs, each consisting of two main procedures: data sampling and policy updating. During data sampling, we simulate the system's evolution under various $q$-profiles generated by the current policy. Specifically, the evolution is divided into $M$ time steps. At each time step, the agent perceives some observables $s_t \in \mathcal{S}$ of the system and selects an action $a_t \in\mathcal{A}$~(which determines $q(t)$ in our study) according to the current policy $\Pi_{\theta}({a_t|s_t})$. In response to the action $a_t$, the system evolves to $s_{t+1}$  and hands over a scalar reward $r_{t}\in\mathcal{R}$ to the agent. After completing the evolution, we compute the cumulative reward $R=\sum_j\gamma^j r_{t_j}$, with $r_{t_j}$ being the reward at the $j$th time step. To account for future rewards, a discount factor $\gamma$ is introduced, which is often set close to $1$ to discourage short-sighted decisions.
Once the data is collected, the agent utilizes it to estimate the value of different state-action pairs ($s_t, a_t$). This includes estimating the advantages, which represent the relative value of actions in each state compared to the expected returns.
The policy $\Pi_{\theta}$ is then updated in order to  maximize the cumulative reward. 
Through iterative optimization of the policy using collected data, estimation of advantages, and updating of policy parameters, the agent is guided towards learning an improved policy that maximizes the cumulative reward. This process persists until the agent attains satisfactory performance or converges to an optimal policy.

\subsubsection{Key elements}

In our study, the system is a spinor atomic condenstate with spin-mixing dynamics governed by Hamiltonian (1) in the main text.
The state space $\mathcal{S}$, action space $\mathcal{A}$, reward function $\mathcal{R}$ and policy function $\Pi_{\theta}$  take the following  meanings.

\begin{itemize}
	\item State space $\mathcal{S}$: 
	A set of four observables, including $\langle \rho_0 \rangle \equiv \langle {N}_0/N \rangle$, $\theta_s \equiv \arg\langle {a}_{+1}^{\dagger} {a}_{-1}^{\dagger} {a}_{0}^2\rangle$, $|{\langle {a}_{+1}^{\dagger} {a}_{-1}^{\dagger} {a}_{0}^2}\rangle|$, and $q(t)/q_{\rm max}$ is chosen to represent the state. Here, the fractional population in $m_F=0$ component, $\langle\rho_0\rangle$, and the spinor phase $\theta_s$ are two variables from the mean field theory for this system, $|\langle {a}_{+1}^{\dagger} {a}_{-1}^{\dagger} {a}_{0}^2\rangle| $ calibrates the first-order quantum correlation for spin-mixing and $q(t)/q_{\rm max}$ denotes the current normalized quadratic Zeeman shift~(QZS).
	Using physically relevant observables like these as state representation facilitates directly generalizable policy due to their independence of $N$ after normalization. Moreover, the final policy can be interpreted with clear physical insights~\cite{guo2021}.
	
	\item Action space $\mathcal{A}$: During the training for small sized systems, the action space consists of the continuous range of the QZS increment with respect to time, $a_j = \Delta q_j/\Delta t_j \in [-2, 2]$, with the time interval at $j$th step $\Delta t_j \in [0, 2\tau/{M}]$, where $M$ denotes the total steps in each trajectory and $\tau$ a guessed total duration.
	The starting point of each sweeping protocol can be manually defined, according to earlier experiences~\cite{guo2021}.
	Based on results from small sized systems, the total duration $\tau$ for subsequent training in larger systems can be predicted by the pre-trained policy.
	The time interval hence is fixed as $\tau/{M}$ for larger sized systems.
	
	\sloppy
	\item Reward function $\mathcal{R}$: For the task of path splitting, two types of object function $f(\psi)$ are chosen as rewards: overlap (or fidelity) between the current and the target states $f(\psi)= {\cal F} = |{\langle\psi(t)|\psi_{\rm tg}\rangle}|^2$ for full quantum simulation in the absence of loss and flipped mean energy $f(\psi)=-\langle\psi(t)|H(q_{\rm tg})|\psi(t)\rangle$ for semi-classical simulation based on truncated Wigner approximation~\cite{steel1998, sinatra2002, hamley2012, liu2021} in the presence of loss. 
	The latter reward is designed to prepare the ground state of $H(q_{\rm tg})$. 
In the recombining task for small sized systems, the metrological performance based on error propagation from $\rho_0$ measurement,
\begin{equation}
	\label{ps_r}
	f(\psi) = |\partial_{\phi}\langle \rho_0 \rangle_{\phi}|/\sqrt{(\Delta \rho_0)_\phi^2 + \sigma_n^2},
\end{equation}
the inverse of phase sensitivity, is employed as reward.  The results from RL for small sized systems reveal the existence of strong correlation between reaching high phase sensitivity and passing through the initial state. Since the calculation of phase sensitivity becomes computationally expensive for large systems, we simplify the recombining training task into preparing the initial state. 
As a result, the reward is set as fidelity $\cal F$ with the initial state (in the absence of loss) or flipped mean energy of the system at large $q$ (with loss).

\item Policy function $\Pi_{\theta}(a|s)$: The policy function maps the state space $\mathcal{S}$ to the action space $\mathcal{A}$. In this work, it assumes a stochastic distribution with a Gaussian form,
\begin{equation}
	\label{policy_fun}
	\Pi_{\theta}(a|s) = \frac{1}{\sqrt{2\pi}\sigma}\exp(-\frac{(a - \mu_{\theta}(s))^2}{2\sigma^2}),
\end{equation}
where $\mu_{\theta}(s)$ is the deterministic policy or the so-called most probable action from the neural network~(NN). 
$\sigma$ is a state-independent parameter representing the standard deviation of noise added on the action $a$.
$\Pi_\theta(a|s)$ is the resulting stochastic policy function used for exploration in the RL training.
When a state is fed into the NN, $\Pi_\theta(a|s)$ determine the rules for choosing an action $a$.

\end{itemize}

\subsubsection{Optimization algorithm: PPO}

We employ the proximal policy optimization (PPO) algorithm~\cite{ppo2017} to update the parameters $\theta$ and search for the optimal policy $\Pi_{\theta}^*$ that maximizes cumulative reward $R$. PPO algorithm is a type of actor-critic method, where the actor network is the policy function and the critic is the value network to evaluate the current state. In our work, residual NNs~\cite{he2015} are adopted to parameterize the actor and the critic,
each containing a residual NN~\cite{he2015} with $[64,32,16,8]$ neurons in the residual block. PPO employs a surrogate objective function to update the policy iteratively. The objective is to maximize the expected reward while ensuring that the policy update does not deviate too much from the previous policy. The surrogate objective function is designed to strike a balance between exploration and exploitation, and it is given by
\begin{equation}
\label{loss_pi}
{\cal L}(\theta) = \min \Big \{\frac{\Pi_{\theta}(a|s)}{\Pi_{\theta_{\rm old}}(a|s)}A(s|a), {\rm clip}\Big(\frac{\Pi_{\theta}(a|s)}{\Pi_{\theta_{\rm old}}(a|s)}, 1-\epsilon, 1+\epsilon \Big)A(s|a) \Big\},
\end{equation}
where $\epsilon$ is a hyperparameter limiting the update rate from the old to the new policy.
$A(s|a)$ is the advantage function estimated by GAE-$\lambda$ method based on the current value $V(s)$ from the critic network, which is updated according to the difference between its output value $V(s)$ and the cumulative reward $R$.
The PPO algorithm employed in this work comes from the OpenAI SpinningUp library (PyTorch version)~\cite{ppo}.
To facilitate the training, we encapsulate the quantum state evolution into a Gym environment as suggested by OpenAI~\cite{gym}.
According to the policy, we choose the action with the maximum probability $\mu_{\theta}^*(s)$ as a deterministic protocol to generate a $q(t)$ profile.

\subsubsection{Discussion on the complexity in RL optimization}
The complexity in RL optimization arises from several factors:
\begin{itemize}
\item Exploration and exploitation trade-off: RL algorithms typically require exploration of different actions to discover optimal policies. This exploration process can be computationally expensive, especially in large action spaces or complex environments. Effectively balancing exploration and exploitation  is crucial for efficient learning. In our study, we use PPO algorithm to incorporate this balance. 

\item Policy representation and function approximation: RL often involves representing policies or value functions using parameterized models such as neural networks. These models introduce additional complexity due to the need for training and optimizing their parameters. The choice of an appropriate policy representation and function approximation method can significantly impact the computational requirements. In our study, we adopted a stochastic distribution with a Gaussian form for the policy function, which presents several advantages. This choice provides flexibility, enabling the policy to adapt to various situations. It also promotes efficient exploration, allowing the RL agent to explore different actions systematically. Moreover, the Gaussian distribution aligns well with the optimization algorithm (PPO) that we used. To approximate the policy function, we employed a parameterized residual neural network architecture. The network consists of a residual block with [64, 32, 16, 8] neurons, allowing for effective learning and representation of complex policies. 

\item Training data collection: RL algorithms typically require interacting with the environment to collect training data. Depending on the complexity of the environment and the number of interactions required, this data collection process can be time-consuming and computationally demanding. To enhance the speed of data collection, significant efforts have been dedicated to improving the efficiency of numerical simulations, including algorithmic optimizations, parallelization and simulation approximation.

\end{itemize}
	
	\section{Transfer learning} \label{tl}
	Due to the almost independence on $N$ of the physical observables in state $s$, training tasks in systems of different $N$ share the same neural network structure, i.e., a larger-sized system inherits the trained neural network from a smaller-sized system as a pre-trained network. 
	This dramatically reduces the required number of training epoches in larger-sized systems, and the total training process becomes efficient since training in larger-sized systems no longer consumes enormous computational resources.
	Such a multi-step training process is adopted for increased system size from $N = 100$ to $N = 5000$ atoms using full quantum simulation of Schr\"odinger equation in the absence of atom loss, and from $N = 5000$ to $N = 10900$ atoms by simulating coupled stochastic diﬀerential equations derived from the quasiprobability distribution based on truncated wigner approximation~(see Sec. \ref{sec2}) with atom loss modeled approximately as one-body decay\cite{guo2021, liu2021}.
	More details of our reinforcement learning~(RL) task, e.g. hyperparameters, PPO algorithms, transfer learning 
	can be found in the Supplemental Material of Ref.~\cite{guo2021}.
	
	\section{Simulating dissipative evolution} \label{sec2}
	We numerically simulate spin mixing dynamics in the presence of atom loss and transverse radio-frequency (RF) magnetic field noise following truncated wigner approximation method presented in Ref.~\cite{liu2021}. The stochastic differential equations to be solved are as follows,
	\begin{equation}
		\left\{
		\begin{aligned}\label{rfsde}
			d\psi_1 &=-ic_2'(t)\left[\psi_0^2\psi_{-1}^*+(|\psi_1|^2-|\psi_{-1}|^2+|\psi_0|^2)\psi_1\right]dt-\frac{\gamma}{2}\psi_1dt+\sqrt{\frac{\gamma}{2}}d\xi_1(t)-\dfrac{i}{\sqrt{2}}d\chi(t)\psi_0,\\
			d\psi_0 &=-ic_2'(t)\left[2\psi_1\psi_{-1}\psi_0^*+\left(|\psi_1|^2+|\psi_{-1}|^2\right)\psi_0\right]dt+iq(t)\psi_0dt-\frac{\gamma}{2}\psi_0dt+\sqrt{\frac{\gamma}{2}}d\xi_0(t)-\dfrac{i}{\sqrt{2}}d\chi(t)\left(\psi_1+\psi_{-1}\right),\\
			d\psi_{-1} &=-ic_2'(t)\left[\psi_0^2\psi_1^*+(|\psi_{-1}|^2-|\psi_1|^2+|\psi_0|^2)\psi_{-1}\right]dt-\frac{\gamma}{2}\psi_{-1}dt+\sqrt{\frac{\gamma}{2}}d\xi_{-1}(t)-\dfrac{i}{\sqrt{2}}d\chi(t)\psi_0,
		\end{aligned}
		\right.
	\end{equation}
	where $c_2'(t)=c_2(t)/N(t)$ with $c_2(t) = c_2e^{-\gamma_c t}$ and $N(t) = Ne^{-\gamma t}$, $d\xi_i(t)$ is complex Wiener noise increment satisfying $\overline{d\xi_i(t)}=0$ and $\overline{d\xi_i^*(t)d\xi_j(t)}=\delta_{i,j}dt$, and $d\chi(t)=\omega_0 d\xi_{\omega}(t)$ denotes Wiener noise increment satisfying $\overline{d\xi_{\omega}(t)}=0$ and $\overline{d\xi_{\omega}(t)^2}=dt$.
	The experiments are simulated with the following parameters: initial atom number $N=10900$, loss rate $\gamma = 0.035~{\rm s}^{-1}$, spin exchange rate $c_2 = -2\pi \times 2.66$~Hz, decay rate $\gamma_c = 0.042~{\rm s}^{-1}$, and RF noise strength $\omega_0 = 2\pi \times 0.006~\sqrt{\rm mHz}$.


	\section{Experimental methods} \label{sec4}
	
	\subsubsection{Initial state preparation}\label{prep}
	In our experiment, a spinor Bose Einstein condensate (BEC) with about $8.3 \times 10^4$ $^{87}$Rb atoms is first produced in the $m_F = -1$ component inside an optical dipole trap formed by two crossed 1064-nm light beams~\cite{luo2017, zou2018}.
	The quantization axis is defined by applying a magnetic field of 815 mG along the direction of gravity, which gives $q_B/|c_2| \simeq 17$ with $q_B$ the quadratic Zeeman shift and $|c_2|$ the spin exchange strength.
	A RF $\pi/2$-pulse is then applied to transfer atoms from $m_F =-1$ to $m_F=0$ followed by ramping up a gradient magnetic field of about 200 G/cm to remove the remaining atoms in the $m_F = \pm 1$ components.
	The number of surviving atoms in the $m_F=0$ component can be flexibly controlled via changing the intensity of the RF pulse.
	Afterwards, the powers of trapping light beams are lowered within 500 ms for further evaporation, while the gradient magnetic field is switched off.
	Next, the trap is held for another 500 ms followed by compression to the final trapping frequencies of $2\pi \times (237, 112,183)$ Hz along three orthogonal directions within 300 ms to produce a condensate of 10900 atoms.
	During the last 300 ms, a red detuned microwave is switched on, maintaining a $q/|c_2| \gg 2$ with suppressed spin mixing dynamics, while the magnetic field is ramped from 815 mG to 537 mG to provide a $q_B/|c_2| \simeq 8$.
	This prepares a BEC sample for subsequent experiment of the RL-designed nonlinear interferometer.
	
	\subsubsection{Measurement of collective spin length}\label{Leff}
	To measure the transverse spin $L_\perp$, a resonant $\pi/2$ RF pulse coupling $\ket{F =1,m_F =0}$ and $\ket{F =1,m_F =\pm1}$ is applied to rotate the probe state before absorption imaging. The collective spin length is calculated according to $\langle {\bf L}^2\rangle = \langle L_x^2 + L_y^2+L_z^2\rangle = 2\langle L_\perp^2\rangle$, where $\langle L_z^2 \rangle=0$ and the rotation symmetry in the $x$-$y$ plane are assumed.

	\subsubsection{Phase encoding}
	For phase encoding, a small rotation along an axis in the $x$--$y$ plane is carried out by a pair of coherent RF pulses. 
	Operationally, this is achieved by first applying a pulse of $12~\mu$s which rotates the state around a given axis by an angle of around 0.1 rad, followed by a companion second pulse of $12~\mu$s with an opposite RF phase rotating the state backward by a similar angle. 
	The slight amplitude difference between the two pulses leads to the intended small angle rotation. 




	\subsubsection{The stability of quadratic Zeeman shift~(QZS)}
	The RL profile is sensitive to fluctuation $\delta q(t)$ of the QZS. 
	For the typical experimental system size ($N\simeq10900$), our simulation shows that fluctuation of $q$ needs to be controlled within the range $\delta q(t)/|c_2|\in [-0.01,0.01]$ so that it has negligible effects on spin-mixing dynamics. 
	To achieve this required stability, we set up feedback-control systems for stabilizing both the microwave power and magnetic field strength. 
	The microwave sensor and magnetic flux gates are both temperature stabilized, in order to prevent drifting detection efficiency. The shot-to-shot fluctuation of our microwave power is around 0.1\% over a 12-hour period, which leads to $\delta q_{\rm MW}/|c_2| \leqslant 0.01$. Our magnetic field control system has a bandwidth of $\sim$1 kHz, which can suppress peak-to-peak field fluctuation to $150~\mu$G, leading to $\delta q_{\rm B}/|c_2|<0.004$. 
	In addition, we compensate for the small magnetic field gradient around BEC with specially designed coils, since the gradient can cause phase separation of spin-up and spin-down components, leading to reduced overlap of different components as well as drift of the spin exchange rate $|c_2|$.

	\subsubsection{Calibrating atom loss} 
	
\begin{figure} [htp!]
		\begin{center}
			\includegraphics[width=0.75\columnwidth]{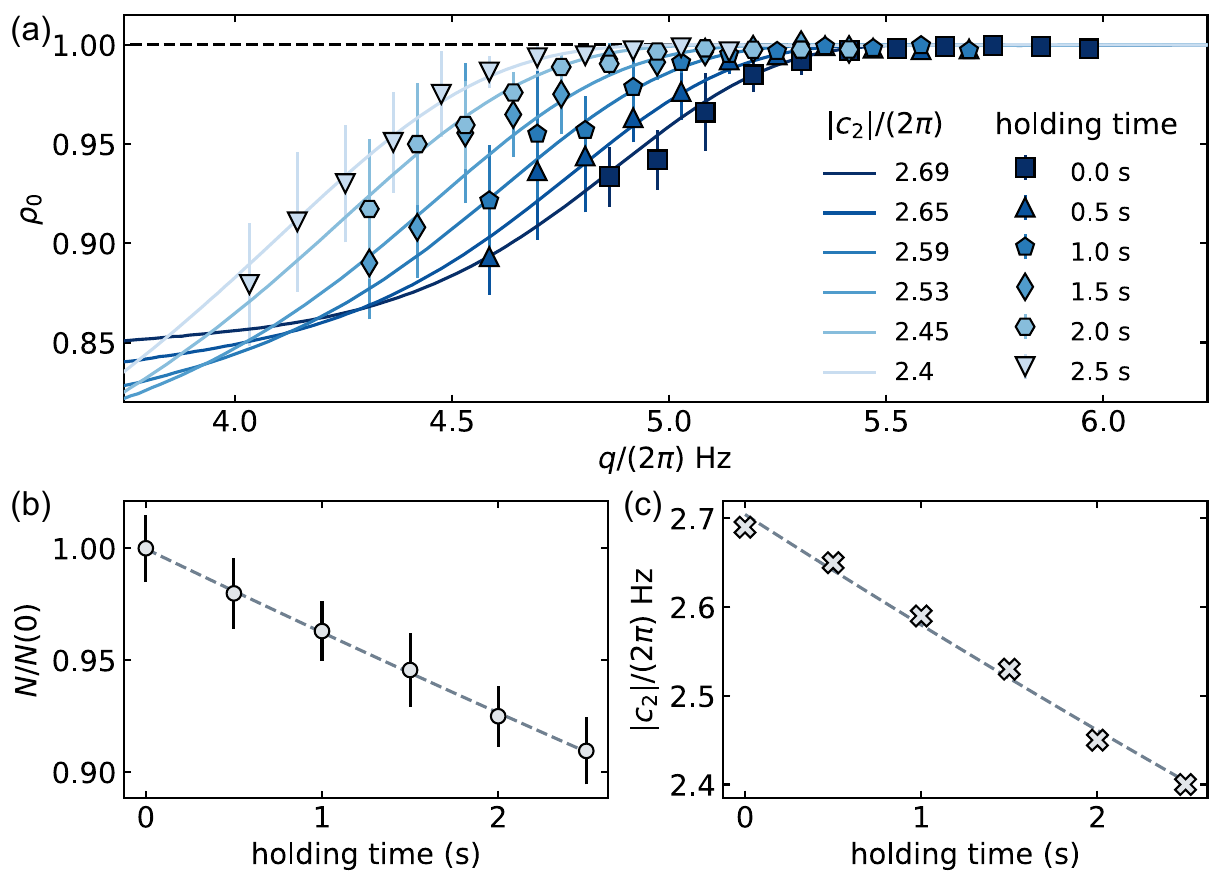}
			\caption{(a) The fractional population, $\rho_0$ after evolving for $400$~ms at different $q$. 
				Different markers correspond to different holding time (at $q/|c_2| \sim 17$) before quench. 
				Each data point comes from 10 continous experimental runs. 
				The solid lines denote numerical results based on truncated Wigner approximation method with the labeled spin exchange rate $c_2$.
				(b) The measured total atom number as a function of holding time. The dashed line denotes an exponentially decaying function with decay rate as a fitting parameter.
				(c) The spin exchange rate $|c_2|$ as a function of holding time.
				The dashed line denotes an exponentially decaying function with decay rate as a fitting parameter.}
			\label{fig_c2decay} 
		\end{center}
	\end{figure}

	We follow Ref.~\cite{hoang2016} to calibrate atom loss rate $\gamma$ and decay rate of $c_2$. We start by holding the condensate with all atoms in $m_F=0$ inside the trap at $q/|c_2| \sim 17$ (such that spin-mixing dynamics is negligible) for different durations (a longer duration leads to more atom loss and smaller $c_2$). 
	This is followed by sudden quench of $q$ to different values and the system is then allowed to evolve for $400$ ms before detecting the remaining atom numbers. 
	The measured results of $\rho_0$ as a function of quenched $q$ are shown in Fig.~\ref{fig_c2decay}(a). Different markers correspond to different holding time before quenching. 
	In each case, the experimental data is compared to numerical results (based on truncated Wigner approximation method) with different $c_2$. 
	The numerical result with the `correct' $c_2$ should match best with the experimental data. 
	This is used to determine the actual value of $c_2$. 
	The numerical results with the calibrated $c_2$ are shown by solid lines in Fig.~\ref{fig_c2decay}(a). 
	By plotting $c_2$ as a function of holding time $t$ and fitting the data with an exponentially decaying function (Fig.~\ref{fig_c2decay}(c)), we extract the decay rate for $c_2$. 
	To calibrate atom loss rate $\gamma$, 
	we plot the total atom number as a function of the holding time and fit the data with exponential decay function (Fig.~\ref{fig_c2decay}(b)). 
	The extracted atom loss rate is found to be $\gamma = 0.035~{\rm s}^{-1}$, and the decay rate $\gamma_c$ of $c_2$ is $1.21\gamma$.
	
	\subsubsection{Calibrating RF noise}
	\begin{figure}[htp!]
		\begin{center}
			\includegraphics[width=0.75\columnwidth]{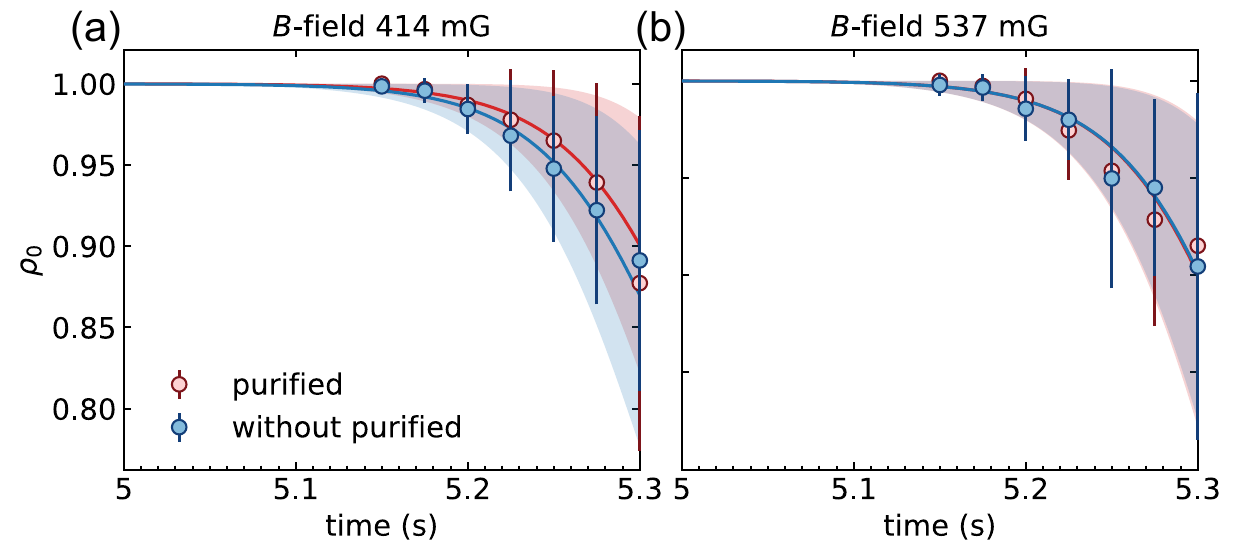}
			\caption{Evolutions of $\rho_0$ at $q=|c_2|$ after holding the condensate with all atoms in $m_F=0$ for $5$~s under $q \gg |c_2|$ (blue circles). 
				The red circles denote the results with atoms in $m_F = \pm 1$ removed before quench. The blue (red) solid line (shaded regions) denotes numerical result with (without) RF noise. (a) and (b) correspond to different bias magnetic fields, with (a) at 414 mG, and (b) at 537 mG.}
			\label{fig_rfnoise} 
		\end{center}
	\end{figure}
	RF noises during the experiments are mainly contributed by the large RF power amplifier and switched-mode power supply, whose influence can be cut off to some degree by inserting additional switches.
	The remaining RF noise exhibits a $1/f$ like spectrum, with most of the noise power concentrated in the low frequency region. 
	The frequency $f$ is propotional to the linear Zeeman shift, hence the bias magnetic field strength.
	The influence of RF noises is thus expected to be smaller at higher magnetic field. 
	This is confirmed in Fig.~\ref{fig_rfnoise}, where we compare the influence of RF noises under two different magnetic fields. 
	In experiment, we first hold the condensate in the polar state with all atoms in $m_F=0$ at $q/|c_2| \gg 2$ for 5 s, which allows RF noise to fully perturb the polar state. 
	The perturbation can be viewed as a small rotation along a random axis in the $L_x$--$L_y$ plane, leading to a small fraction of atoms transferred to the $m_F = \pm 1$ components.
	This signal can be further amplified by allowing the system evolve at $q/|c_2|=1$ for a short duration. 
	The strength of RF noise can be inferred from comparing measured evolution of $\rho_0$ with simulated ones using different noise strength. 
	The numerical result with the `correct' RF noise strength should match the best with experimental data, which calibrates the strength of RF noise. 
	For a magnetic field of 414 mG, the inferred RF noise is $\omega_0 = 2\pi\times 0.012~\sqrt{\rm mHz}$ (Fig. \ref{fig_rfnoise}(a)), while results at 537 mG imply a RF noise $\lesssim 2\pi\times 0.006~\sqrt{\rm mHz}$ (Fig. \ref{fig_rfnoise}(b)), smaller than that at 414 mG, as expected.
	Therefore, in order to reduce the influence of RF noise, higher bias magnetic field is preferred. 
	The required control precision over QZS, on the other hand, appeals for low magnetic field. 
	As a compromise between RF noise and instability of QZS, we choose to perform experiments at the highest magnetic field (537 mG) that still guarantees the stability of $q$ with fluctuation $\delta q(t)/|c_2|\in [-0.01,0.01]$.

	\subsubsection{Calibrating the RF rotation angle}
\begin{figure}[htp!]
		\begin{center}
			\includegraphics[width=0.65\columnwidth]{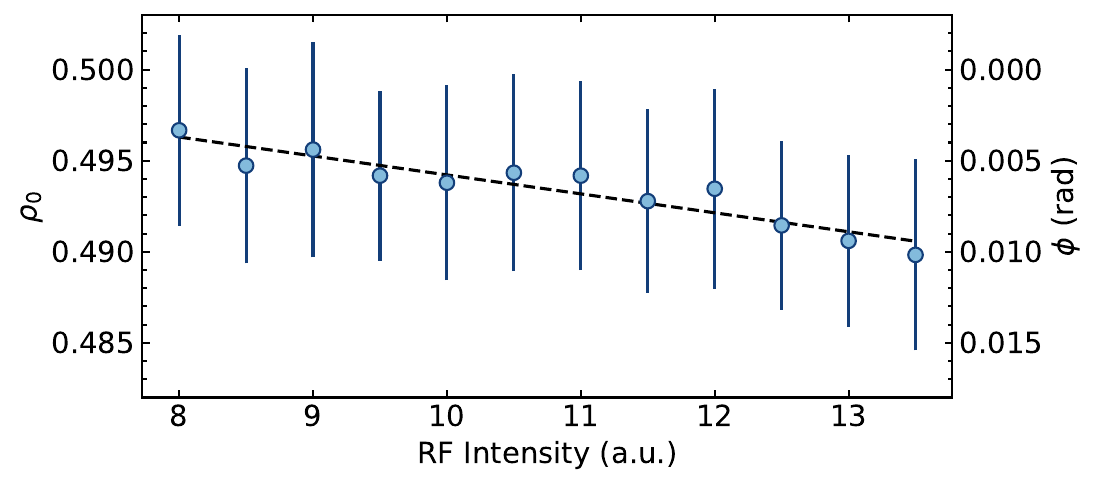}
			\caption{Measured $\rho_0$ under different RF intensity. Each data comes from 50 experimental repetitions. The dashed line represents a linear fitting.}
			\label{fig_rfcali} 
		\end{center}
	\end{figure}

	To calibrate the (small) angle of RF rotation for phase encoding, we first apply a RF $\pi/4$-pulse to transform the polar state to a coherent state $(-i\ket{-1}/2 + \ket{0}/\sqrt{2} - i\ket{+1}/2)^{\otimes N}$ with $\langle \rho_0 \rangle = 1/2$. Then a small angle RF rotation composed of two RF pulses is applied to the coherent state. The population in $m_F=0$ decreases with the small rotation angle $\phi$ approximately as $\langle \rho_0 \rangle \simeq 1/2 -\phi$. By fitting the data of $\rho_0$ for different RF intensity (which is proportional to the rotation angle) with a linear function, a one-to-one correspondence between RF intensity and rotation angle is calibrated~(Fig. \ref{fig_rfcali}).
	
	\subsubsection{Calibrating atom number detection}
	\begin{figure}[htp!]
		\begin{center}
			\includegraphics[width=0.6\columnwidth]{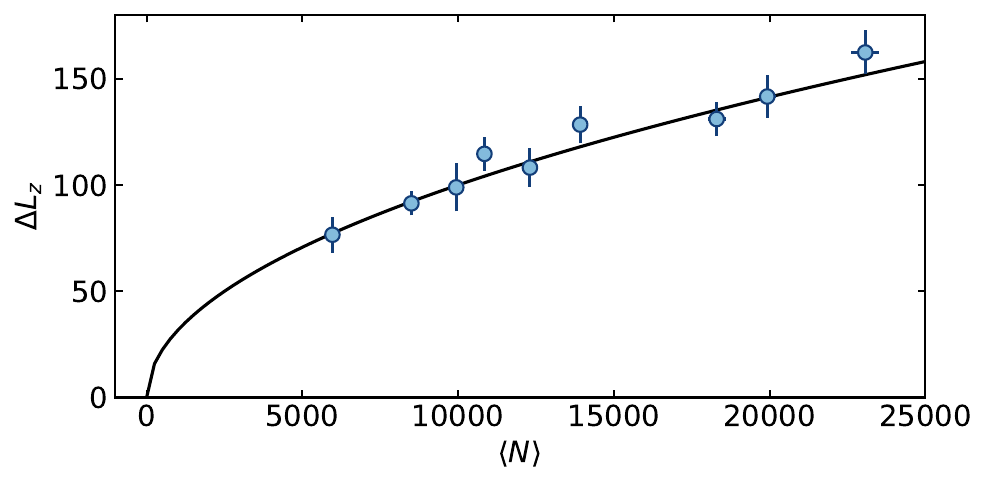}
			\caption{Measured $\Delta L_z$ vs $N$, compared with the projection noise $\sqrt{N}$ (black solid line).}
			\label{fig_caliN} 
		\end{center}
	\end{figure}	
	To check the accuracy of atom number detection, we prepare atomic coherent state $((\ket{1}+\ket {-1})/\sqrt{2})^{\otimes N}$ by applying a RF $\pi/2$-pulse to the polar state $|0\rangle^{\otimes N}$, and then measure the variance of
	atom number difference between the $m_F =\pm 1$ components, $\Delta L_z = \Delta(N_1 - N_{-1})$. In Fig. \ref{fig_caliN}, we find the measured $\Delta L_z$ agrees well with uncorrelated quantum projection noise (black solid line), which confirms the atom number detection accuracy.

	\section{More data} \label{sec5}
	\subsubsection{Training results in the presence of experimental imperfection}
\begin{figure}[!htp]
		\begin{center}
			\includegraphics[width=0.99\columnwidth]{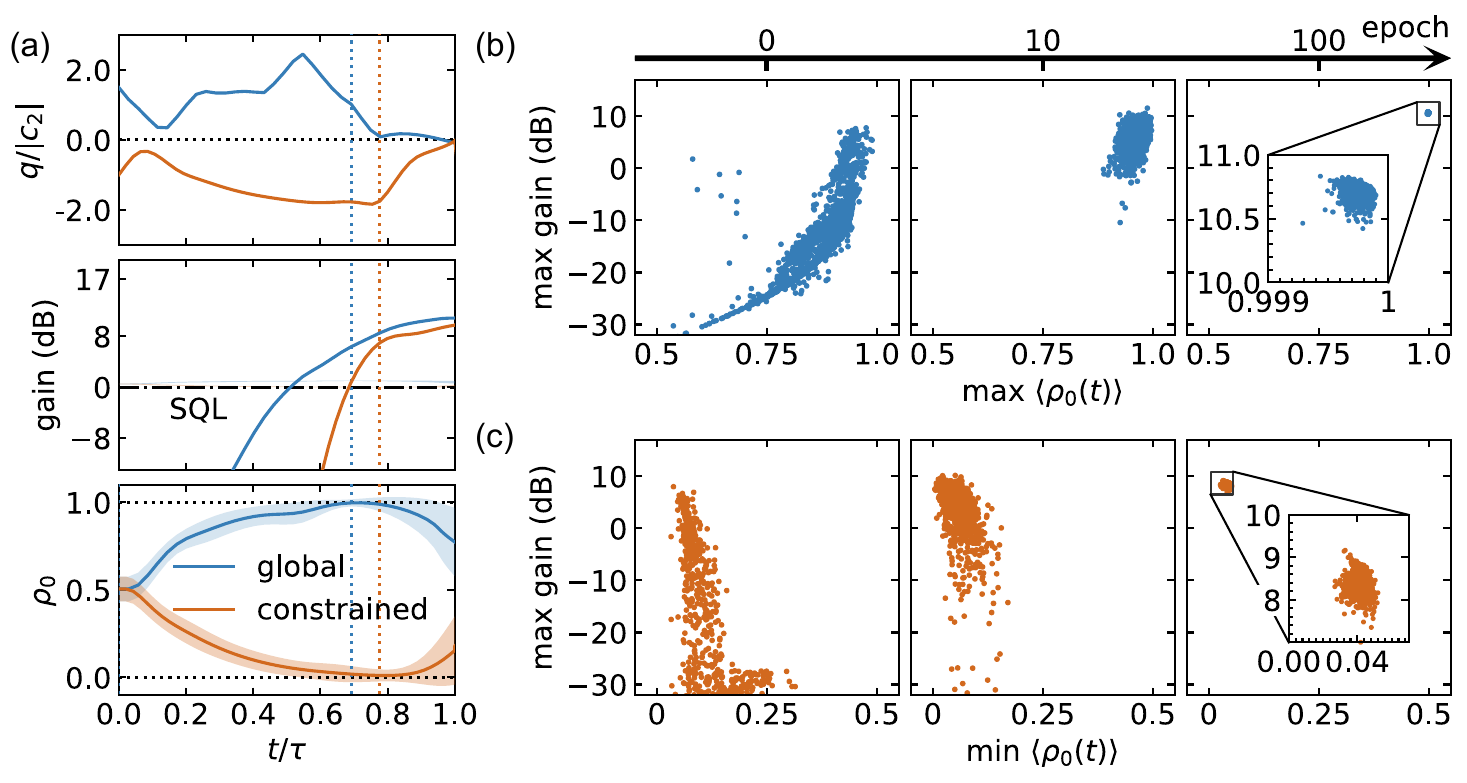}
			\caption{(a) RL training results for the readout operation in a small system of $N = 50$ particles with $\sigma_n=0.02$. The colors of data represent the same meaning as Fig.~2 in the main text.
				The blue (orange) vertical dotted lines mark the temporal instant of reaching maximal (minimal) $\rho_0$ when noise is included.
				(b) Maximal achievable metrological gain vs maximal accessible $\rho_0$ during the readout
				process from 2000 trajectories for each sampled according to the policy at 0, 10 and 100 epoch. 
				(c) Similar results as (b), but for the case with the constrain $q < 0$.}
			\label{fig_50moredata} 
		\end{center}
	\end{figure}

	In the presence of noise due to experimental imperfections, which is modeled by augmenting a constant variance $\sigma_n^2$ to $(\Delta \rho_0)^2_{\phi}$, the optimal strategy found [Fig.~\ref{fig_50moredata}(a) lower panel] again shows the system passes through the unstable fix points as the Fig.~2 of the main text.
	The policy from global searching~(with no constrain on $q$) drives the system towards the polar state (fixed point on the north pole), while the one with the constrain $q<0$ drives it towards the Twin-Fock state (fixed point on the south pole).
	After passing through the unstable fixed points, the system ends up at a final state with large $(\Delta \rho_0)^2$. 
	This latter stage can be interpreted as a compromise to submerge the effect of noise by parametric amplification~\cite{arlt2010} which magnifies both signal and quantum noise.
	The scatter plots of learning process also demonstrate the importance of passing through unstable fixed points in achieving high phase sensitivity [Fig.~\ref{fig_50moredata}(b) and (c)].
	
	\subsubsection{Probe state preparation}
\begin{figure}[!htp]
		\begin{center}
			\includegraphics[width=0.99\columnwidth]{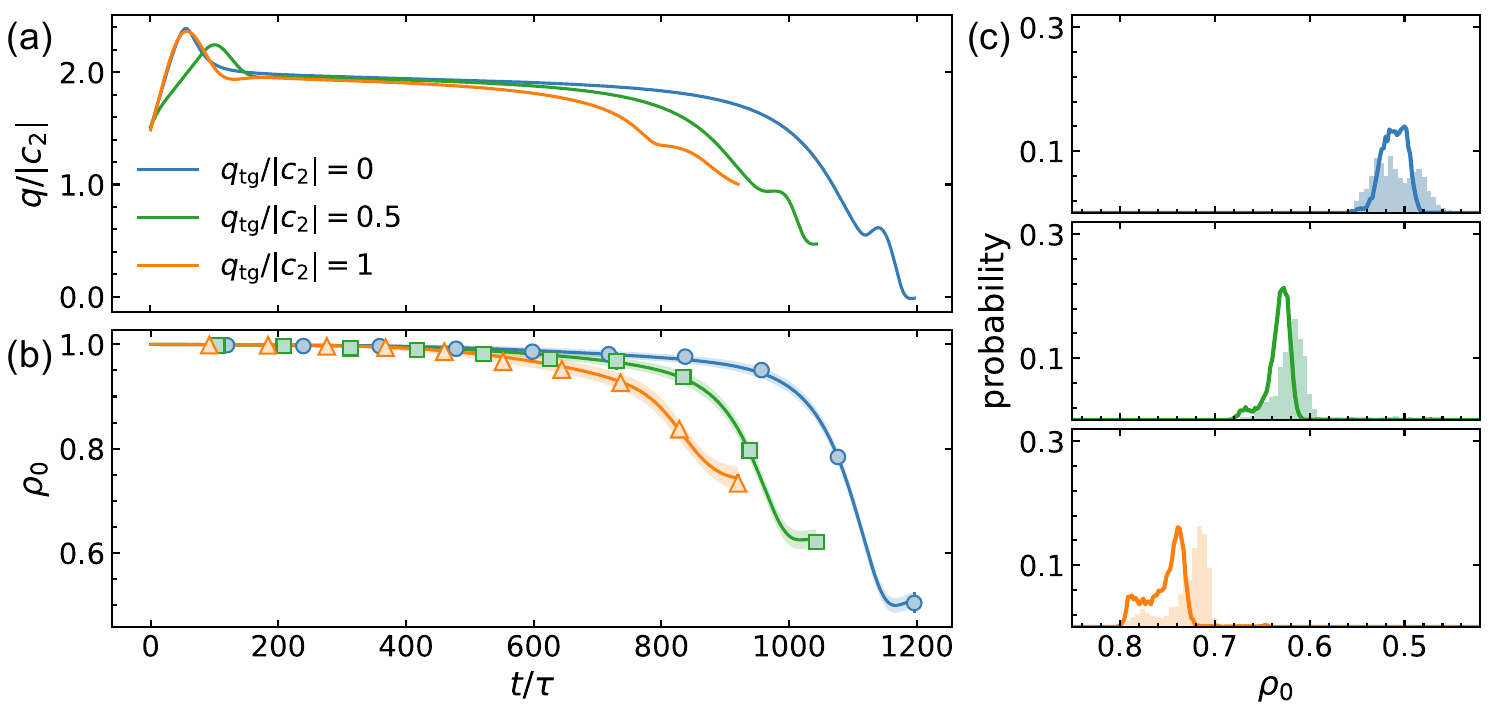}
			\caption{(a) RL proﬁles for different probe states with $q_{\rm tg}/|c_2| = 0$ (blue), 0.5 (green), and 1 (orange), starting from the same initial polar state with $N \simeq 10900$ atoms in $m_F=0$. 
				(b) The corresponding evolutions of $\rho_0 $, with lines (shaded regions) denoting simulated mean values (uncertainties). Data points from 10 (100) experimental runs for the state preparation process are marked by circles, squares, or triangles, respectively for probe states with $q_{\rm tg}/|c_2|=$ 0, 0.5, and, 1.
				(c) Measured distributions for the three probe states with $q_{\rm tg}/|c_2| =$ 0 (blue), 0.5 (green), and 1 (orange) in $\rho_0$ based on 500 continuous experimental runs. The solid lines denote numerical simulation results.
			}
			\label{fig_evodata} 
		\end{center}
	\end{figure}

	Figure~\ref{fig_evodata}(a) shows the $q$ profiles for preparing the three probe states discussed in the main text, whose distributions in $\rho_0$ are shown in Fig.~\ref{fig_evodata}(c). 
	The measured evolutions of fractional population for the $m_F=0$ component during the state preparation process are shown in Fig.~\ref{fig_evodata}(b).
	
	\section{Spin-1 balanced Dicke state} \label{sec:dicke}
	In this section, we give a brief introduction to spin-1 balanced Dicke state $|L=N, m=0\rangle$. It is the common eigenstate  of the collective spin operators ${\bf L}^2$ and $L_z$, with respective eigenvalues $N(N+1)$ and $0$. Here, ${\bf L} \equiv (L_x,L_y,L_z)$, $L_k = \sum_j{F^{(j)}_k}$ with $F^{(j)}_k$ being the spin operator of the $j$th spin-1 particle along the $k$~($=x,y,z$)~direction. 
	The Hamiltonian of the spin-1 condensate discussed in the main text can be written as $H=c_2/(2N){\bf L}^2 - qN_0$ with $c_2<0$. Therefore, the spin-1 balanced Dicke state $|L=N, m=0\rangle$ is the ground state of the system at $q=0$.
	
	\begin{figure}[!htp]
		\begin{center}
			\includegraphics[width=0.25\columnwidth]{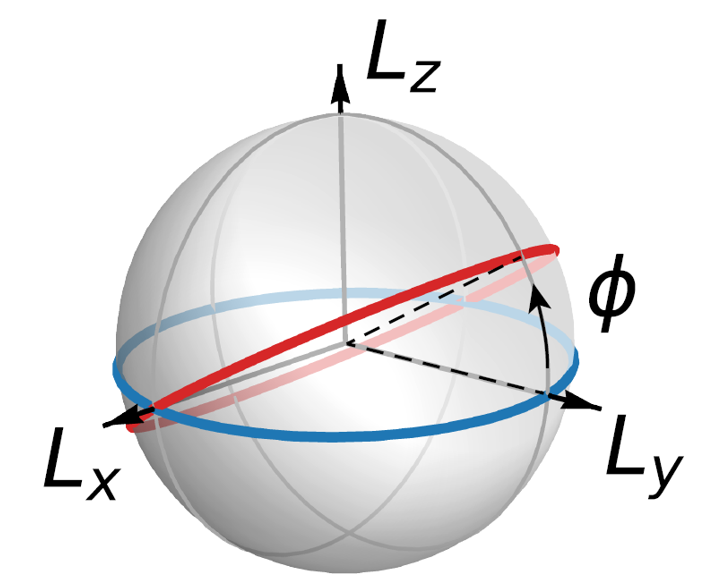}
			\caption{Bloch sphere representation of spin-1 balanced Dicke state~(blue band) and the state rotated from it around $L_x$ by angle $\phi$ (red band). 
			}
			\label{dicke} 
		\end{center}
	\end{figure}
	
	As the balanced Dicke state is an eigenstate of $L_z$ with eigenvalue $0$, the direction of its transverse spin is totally indeterminate according to the Heisenberg uncertainty principle. Hence, it can be represented as an annulus on the generalized Bloch sphere of radius $\sqrt{N(N+1)}$, as shown by the blue band in Fig.~\ref{dicke}.

	The Dicke state is entangled in the spin degrees of freedom among the constituent particles. For illustration, let's take the two-particle case as an example, whose balanced Dicke state can be written as
	\begin{equation}
		|L=2,m=0\rangle = \frac{1}{\sqrt{6}}[|m_1=1\rangle \otimes |m_2=-1\rangle + 2|m_1=0\rangle \otimes |m_2=0\rangle +|m_1=-1\rangle \otimes |m_2=1\rangle].
	\end{equation}
	Due to the fact that the total magnetization $m=m_1+m_2$ is equal to zero, the spin states of the two particles are entangled. Namely, if the first particle is in spin state $|m_1=1\rangle$, the second particle must be in state $|m_2=-1\rangle$, and so on.
	
	The entangled Dicke state can provide a high phase sensitivity approaching the Heisenberg limit. To illustrate it, let's consider a phase encoding process which rotates the Dicke state around $L_x$ by angle $\phi$. The rotated state is represented by the red band in Fig.~\ref{dicke}. One can see that a larger rotation angle gives rise to a stronger fluctuation of $L_z$. Therefore, one can extract the phase by measuring $\langle L_z^2 \rangle$. It can be shown that the corresponding phase sensitivity according to error-propagation theory is given by
	\begin{equation}
		\Delta \phi = \frac{1}{\sqrt{2N(N+1)}},
	\end{equation}
	which approaches the three-mode Heisenberg limit $1/(2N)$.
	
	Such a direct detection approach, however, is fragile to detection noise, whose presence reduces the phase sensitivity approximately to~\cite{zou2018}
	\begin{equation}
		\Delta \phi = \sqrt{\frac{3\sigma_n^2+1/2}{N(N+1)}},
	\end{equation}	
	where $\sigma_n$ denotes detection noise, typically difficult to reach below  the order of $\sqrt{N}$. Returning to the Bloch sphere representation, the impact of detection noise can be understood as widening or broadening the band associated with the quantum states. This widening makes it challenging to distinguish between the two states, namely the unrotated and rotated states. In the main text, we use RL to find the optimal approach for modulating $q(t)$ to implement a detection-noise-robust nonlinear readout process, thereby inducing significant distinctions in the two resulting final states.
	
	\section{Comparison with time-reversal protocol} \label{sec6}
	
\begin{figure}[!htp]
		\begin{center}
			\includegraphics[width=0.99\textwidth]{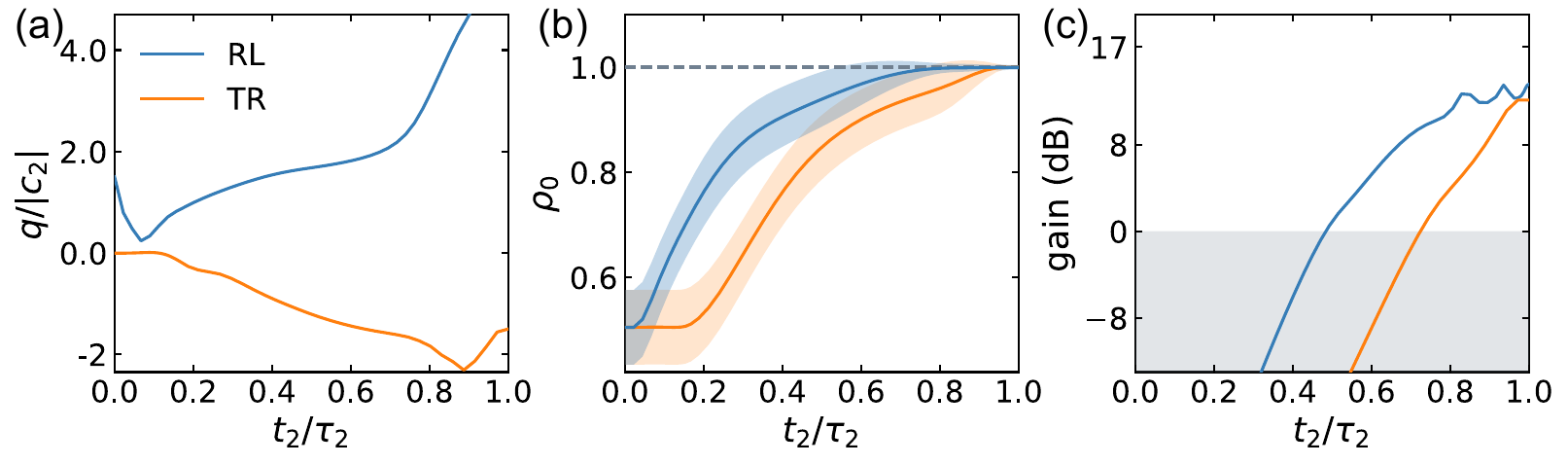}
			\caption{(a) The optimal profiles for ramping $q(t)$ in the nonlinear readout operation respectively for RL (blue) and TR protocols (orange) at $N = 50$. (b) and (c) are the corresponding evolutions for $\rho_0$ and metrological gain.}
			\label{fig_50tr} 
		\end{center}
	\end{figure}

\begin{figure}[!htp]
		\begin{center}
			\includegraphics[width=0.99\textwidth]{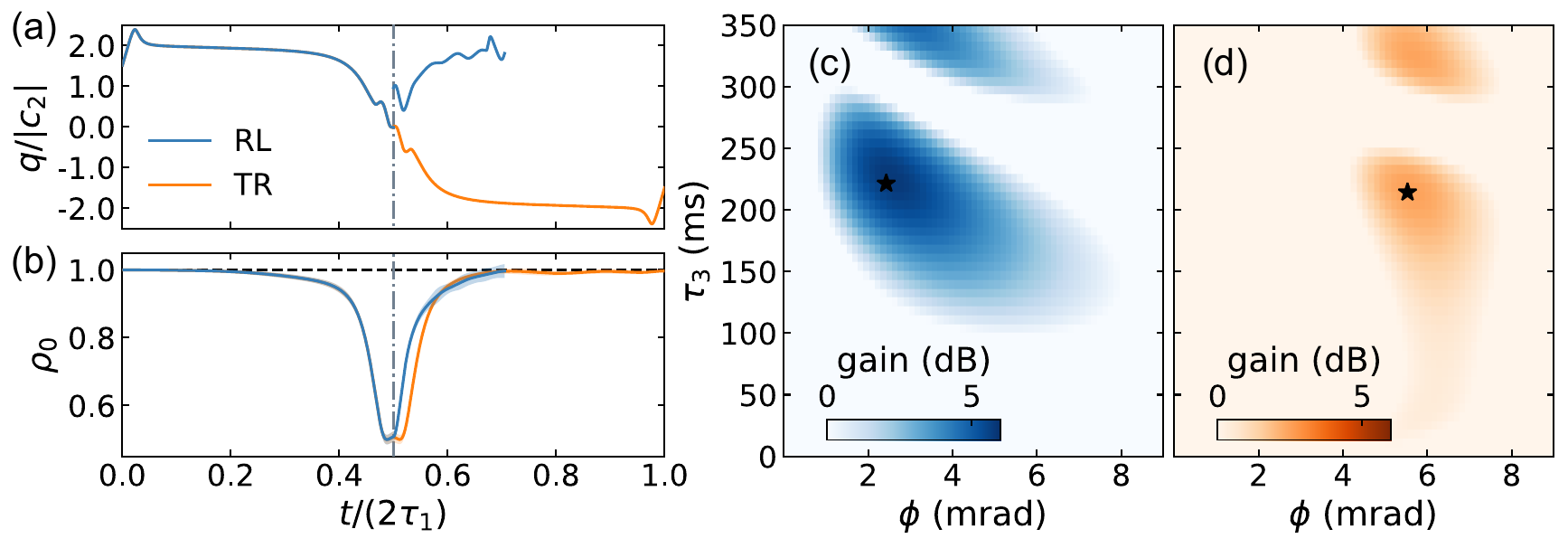}
			\caption{(a) The optimal profiles for ramping $q(t)$ in the entanglement generation process~(left of the vertical dot-dashed line) and in the nonlinear readout process~(right of the vertical dot-dashed line) for the RL (blue) and TR (orange) protocols at $N =10900$ with $\tau_1 = 1196$~ms. 
				(b) The corresponding simulated evolutions of $\rho_0$. (c) and (d) are the simulated dependence of  metrological performance in the vicinity of $\phi=0$ on the duration $\tau_3$ of spin mixing dynamics after the system passes through the initial state. The maximal gain of RL protocol is 5.8~dB, while the TR protocol achieves only 2.54~dB [respectively marked by stars in (c) and (d)].}
			\label{fig_exptr} 
		\end{center}
	\end{figure}

	In this section, we compare our protocol from RL to the time-reversal (TR) protocol with both the signs of the interaction and the control field flipped during the recombining process.
	For a small system with $N = 50$ in the absence of atom loss, the performances of RL and TR are found comparable, as shown in Fig. \ref{fig_50tr}. 
	This scenario changes when it comes to experimental sized system subjected to noise and loss (Fig. \ref{fig_exptr}). The duration for the RL nonlinear readout operation is limited to $\tau_2 \lesssim \tau_1/2$ in order to mitigate the influence of atom loss and RF noise. 
	In comparison, the TR protocol suffers from more loss and noise because $\tau_2=\tau_1$. 
	As shown in Fig. \ref{fig_exptr}(c) and (d), the RL protocol outperforms the TR protocol in this case.
	
	\section{Nonlinear readout in a spin-1/2 system}	\label{spin-half}
	In this section, we present results from additional studies on nonlinear readout of entangled states in a spin-1/2 system. 
	
	The system is described by the following Hamiltonian
	\begin{equation}\label{HOAT}
		H = \chi J_z^2 + \Omega J_x.
	\end{equation}
	Here, $J_\nu = \sum_k{\sigma_{\nu}^{(k)}/2}$~($\nu= x,y,z$) denotes collective spin components with the Pauli matrices $\sigma_{x,y,z}^{(k)}$ for the $k$th particle. The first term in the Hamiltonian~\eqref{HOAT} describes the one-axis twisting interaction of strength $\chi$, and the second term denotes a transverse field of strength $\Omega$.
	
	We generate the entangled probe state $|\psi\rangle$ by evolving the system from a coherent spin state with all the spins pointing to the $x$-direction under the system Hamiltonian~\eqref{HOAT} with $\chi=\Omega=1$ for $\tau=0.15$ and $N=100$. The generated state exhibits a large spin fluctuation along the $y$-axis. The phase is encoded by rotating the probe state by an angle $\phi$ along the $y$-axis, i.e., the phase encoding process is described by evolution operator $U_\phi=e^{-i\phi J_y}$. Our goal is to find a detection-noise-robust strategy for distinguishing the prepared state with and without phase perturbation, i.e., $|\psi\rangle$ and $U_\phi|\psi\rangle$. In the traditional time-reversal~(TR) approach, one lets the system evolve under $-H$ for the same duration of time $\tau$ before measurement. This nonlinear readout process leads to significant distinctions in the resulting final states, and thus promises high phase sensitivity even in the presence of detection noise. 
	
	\subsubsection{RL results}
	
	Here, we employ RL to find the optimal approach for modulating the transverse field $\Omega$ to implement a nonlinear readout process. 
	In this case, the state space encompasses four observable quantities: $\langle J_x\rangle/j$, $\langle J_x^2\rangle/j^2$, $\langle J_y^2\rangle/j^2$, and $\langle J_y J_z + J_z J_y\rangle/j^2$, where $j=N/2$. Note that the state space does not encompass all the terms up to the second order. This omission is justified by the fact that the following conditions hold: $\langle J_y\rangle=\langle J_z\rangle=\langle J_x J_z + J_z J_x\rangle=\langle J_y J_x + J_x J_y\rangle=0$. Additionally, $\langle J_z^2\rangle = j(j+1)-\langle J_x^2\rangle - \langle J_y^2\rangle$. The action space is set to be $\Omega(t)$, the strength of the transverse field. The metrological performance based on error propagation from $J_y$ measurement,
	\begin{equation}
		\label{ps_r}
		(\Delta \phi)^{-1} = |\partial_{\phi}\langle J_y \rangle_{\phi}|/(\Delta J_y)_\phi,
	\end{equation}
	the inverse of phase sensitivity, is employed as reward.

\begin{figure}[!htp]
		\begin{center}
			\includegraphics[width=0.99\textwidth]{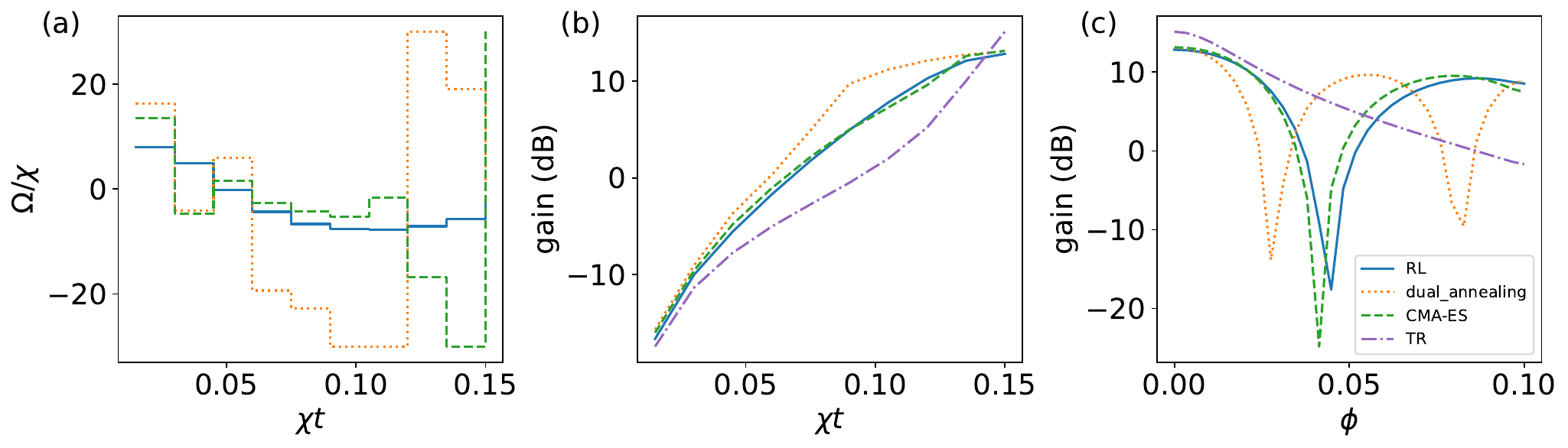}
			\caption{Nonlinear readout of an entangled state generated by evolving the system from a coherent spin state with all the spins pointing to the $x$-direction under the system Hamiltonian~\eqref{HOAT} with $\chi=\Omega=1$ for $\tau=0.15$ at $N=100$. (a) The optimal strategy of ramping the control field $\Omega(t)$ for the nonlinear readout. The total evolution is split into $10$ time steps. Within each time step, the control field is a constant. (b) The evolution of the optimal metrological gain $-20\log_{10}{(\Delta\phi/\Delta\phi_{\rm SQL})}$ during the nonlinear readout process, over the standard quantum limit $\Delta \phi_{\rm SQL}=1/\sqrt{N}$. (c) The metrological gain at small phases for the state at the end of the nonlinear readout process. Different curves correspond to different methods. The blue solid line denotes the results from RL. The orange dotted line denotes the results from stimulated annealing. The green dashed line denotes the results from CMA-ES. The purple dot-dashed line denotes the results from time-reversal protocol, where the nonlinear readout is performed by evolving the system under $-H$ for $\tau$. 
			}
			\label{fig_spin_half} 
		\end{center}
	\end{figure}

	The results obtained from RL are illustrated in Fig.~\ref{fig_spin_half} using blue solid lines. In panel (a), the optimal strategy for modulating the control field $\Omega(t)$ during the nonlinear readout is presented. Panel (b) displays the evolution of the optimal metrological gain $-20\log_{10}{(\Delta\phi/\Delta\phi_{\rm SQL})}$ throughout the nonlinear readout process. The optimal metrological gain corresponds to the maximal gain within a small phase range of $\phi \in [0,0.1]$, where the standard quantum limit $\Delta \phi_{\rm SQL}=1/\sqrt{N}$.
	Panel (c) provides an overview of the metrological gain achieved at the end of the nonlinear readout process over the phase range of $[0,0.1]$. It can be seen that, in this case, the optimal metrological gain is attained when $\phi=0$. Namely, the optimal metrological gain for $\chi t=0.15$ in panel (b) corresponds to the metrological gain at $\phi=0$, which is depicted in panel (c).
	As a comparison, we also show the results from traditional TR approach, as depicted by purple dot-dashed line. From (b) one can see that the TR approach outperforms towards the end. While for the majority of the nonlinear readout process, RL prevails over TR. This confers a significant advantage to RL approach when dissipation or noise are taken into account as in realistic application scenarios. 
	
	\begin{figure}[!htp]
		\begin{center}
			\includegraphics[width=0.99\textwidth]{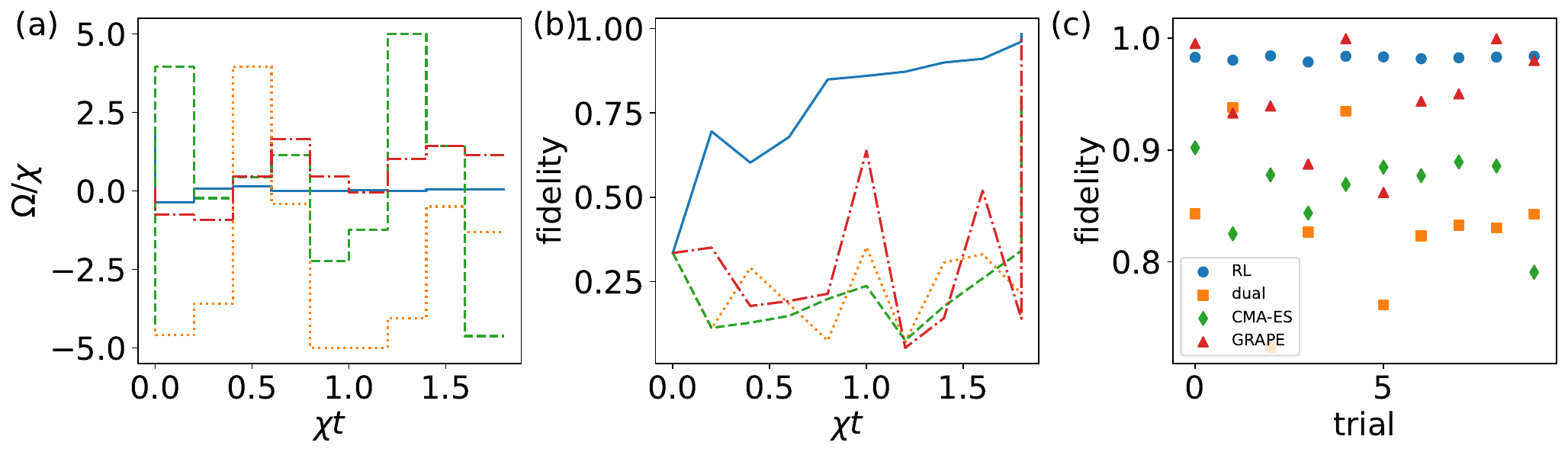}
			\caption{State preparation targeted at twin-Fock state by evolving the system from a coherent spin state with all the spins pointing to the $x$-direction under the system Hamiltonian~\eqref{HOAT} with $\Omega(t)$ as the control parameter for $\chi\tau=2$ and $N=50$. (a) The optimal strategy of ramping the control field $\Omega(t)$ for the state preparation. The total evolution is split into $10$ time steps. Within each time step, the control field is a constant. (b) The evolution of fidelity between the instantaneous state and the target state, $|\langle \psi_{\rm target}|\psi(t)\rangle|$, during the preparation process. (c) The final fidelity for different trials. Different curves correspond to different methods. The blue solid line denotes the results from RL. The orange dotted line denotes the results from stimulated annealing. The green dashed line denotes the results from CMA-ES. The red dot-dashed line denotes the results from GRAPE. 
			}
			\label{fig_tf} 
		\end{center}
	\end{figure}
	
	\subsubsection{A comparison between RL and traditional optimization techniques }
	
	We also showcase the outcomes obtained by employing traditional black-box optimization techniques, such as simulated annealing (depicted by the orange dotted line in Fig.~\ref{fig_spin_half}) and CMA-ES~(Covariance Matrix Adaptation Evolution Strategy)~(represented by the green dashed line in Fig.~\ref{fig_spin_half}). To implement simulated annealing, we utilize the `dual\_annealing' function from the `scipy.optimize' library. For CMA-ES, we employ the `fmin' function from the `cma' Python library. In both cases, the objective is to maximize the metrological gain $G$, i.e., minimizing $-G$. The optimization is bounded within the range of $\Omega/\chi \in [-30, 30]$. The optimal control fields obtained from these two methods are not as smooth as that from RL, as shown in (a), although for the metrological performances, all the methods yield comparable results. 
	
	The performance of these algorithms can vary based on specific characteristics of the problem.
	As an example, we consider the process of preparing a target state, the so called twin-Fock state~\cite{holland1993}. This state exhibits a high level of entanglement, with the particles equally distributed into two modes, and offers a remarkable phase sensitivity close to the Heisenberg limit. The main objective in this scenario is to maximize the fidelity between the prepared state and the target state. The findings obtained from various optimization techniques are summarized in Fig.~\ref{fig_tf}. It is evident that in this particular case, RL outperforms both simulated annealing and CMA-ES. Additionally, we present results obtained using GRAPE (Gradient Ascent Pulse Engineering), a method aimed to discover control pulses that effectively guide the quantum system towards a desired target state or accomplish a specific quantum operation with high fidelity. For GRAPE optimization, we employ the `optimize\_pulse\_unitary' function from QuTip~\cite{JOHANSSON20131234}. As expected, GRAPE achieves the highest fidelity at the end of the evolution, while its performance during the process is not as good as RL. It is worth noting that GRAPE often requires a well-defined and well-posed initial guess for the control pulses, which can influence the final results. Figure~\ref{fig_tf}(c) displays the ultimate fidelity achieved in various trials. It is apparent that the performance of GRAPE exhibits fluctuations, whereas RL demonstrates remarkable stability. 
	When compared to RL, GRAPE can be considered relatively constrained in flexibility as it typically relies on gradient-based optimization algorithms to iteratively refine the control pulses, aiming to maximize the fidelity of the quantum system. Therefore, the application of GRAPE encounters limitations when confronted with problems that lack an available or accessible gradient function, as in previous example. In comparison, RL has the advantage of being able to adapt and learn control policies in a more flexible and data-driven manner, which becomes beneficial in situations where the optimal control strategy is not well-defined or difficult to determine a priori.
	RL possesses another notable advantage in its remarkable generalization capabilities, which allows for transfer learning across diverse scenarios. By transferring the acquired knowledge in smaller sized systems, RL accelerates the training process for  larger systems. This utilization of pre-trained networks yields significant benefits in optimizing performance for complex problems.


\begin{thebibliography}{61}%
\makeatletter
\providecommand \@ifxundefined [1]{%
 \@ifx{#1\undefined}
}%
\providecommand \@ifnum [1]{%
 \ifnum #1\expandafter \@firstoftwo
 \else \expandafter \@secondoftwo
 \fi
}%
\providecommand \@ifx [1]{%
 \ifx #1\expandafter \@firstoftwo
 \else \expandafter \@secondoftwo
 \fi
}%
\providecommand \natexlab [1]{#1}%
\providecommand \enquote  [1]{``#1''}%
\providecommand \bibnamefont  [1]{#1}%
\providecommand \bibfnamefont [1]{#1}%
\providecommand \citenamefont [1]{#1}%
\providecommand \href@noop [0]{\@secondoftwo}%
\providecommand \href [0]{\begingroup \@sanitize@url \@href}%
\providecommand \@href[1]{\@@startlink{#1}\@@href}%
\providecommand \@@href[1]{\endgroup#1\@@endlink}%
\providecommand \@sanitize@url [0]{\catcode `\\12\catcode `\$12\catcode
  `\&12\catcode `\#12\catcode `\^12\catcode `\_12\catcode `\%12\relax}%
\providecommand \@@startlink[1]{}%
\providecommand \@@endlink[0]{}%
\providecommand \url  [0]{\begingroup\@sanitize@url \@url }%
\providecommand \@url [1]{\endgroup\@href {#1}{\urlprefix }}%
\providecommand \urlprefix  [0]{URL }%
\providecommand \Eprint [0]{\href }%
\providecommand \doibase [0]{http://dx.doi.org/}%
\providecommand \selectlanguage [0]{\@gobble}%
\providecommand \bibinfo  [0]{\@secondoftwo}%
\providecommand \bibfield  [0]{\@secondoftwo}%
\providecommand \translation [1]{[#1]}%
\providecommand \BibitemOpen [0]{}%
\providecommand \bibitemStop [0]{}%
\providecommand \bibitemNoStop [0]{.\EOS\space}%
\providecommand \EOS [0]{\spacefactor3000\relax}%
\providecommand \BibitemShut  [1]{\csname bibitem#1\endcsname}%
\let\auto@bib@innerbib\@empty
\bibitem [{\citenamefont {Giovannetti}\ \emph {et~al.}(2004)\citenamefont
  {Giovannetti}, \citenamefont {Lloyd},\ and\ \citenamefont
  {Maccone}}]{giovannetti2004}%
  \BibitemOpen
  \bibfield  {author} {\bibinfo {author} {\bibfnamefont {Vittorio}\
  \bibnamefont {Giovannetti}}, \bibinfo {author} {\bibfnamefont {Seth}\
  \bibnamefont {Lloyd}}, \ and\ \bibinfo {author} {\bibfnamefont {Lorenzo}\
  \bibnamefont {Maccone}},\ }\bibfield  {title} {\enquote {\bibinfo {title}
  {Quantum-enhanced measurements: Beating the standard quantum limit},}\ }\href
  {https://www.science.org/doi/abs/10.1126/science.1104149} {\bibfield
  {journal} {\bibinfo  {journal} {Science}\ }\textbf {\bibinfo {volume}
  {306}},\ \bibinfo {pages} {1330--1336} (\bibinfo {year} {2004})}\BibitemShut
  {NoStop}%
\bibitem [{\citenamefont {Pezz\`e}\ \emph {et~al.}(2018)\citenamefont
  {Pezz\`e}, \citenamefont {Smerzi}, \citenamefont {Oberthaler}, \citenamefont
  {Schmied},\ and\ \citenamefont {Treutlein}}]{luca2018}%
  \BibitemOpen
  \bibfield  {author} {\bibinfo {author} {\bibfnamefont {Luca}\ \bibnamefont
  {Pezz\`e}}, \bibinfo {author} {\bibfnamefont {Augusto}\ \bibnamefont
  {Smerzi}}, \bibinfo {author} {\bibfnamefont {Markus~K.}\ \bibnamefont
  {Oberthaler}}, \bibinfo {author} {\bibfnamefont {Roman}\ \bibnamefont
  {Schmied}}, \ and\ \bibinfo {author} {\bibfnamefont {Philipp}\ \bibnamefont
  {Treutlein}},\ }\bibfield  {title} {\enquote {\bibinfo {title} {Quantum
  metrology with nonclassical states of atomic ensembles},}\ }\href {\doibase
  10.1103/RevModPhys.90.035005} {\bibfield  {journal} {\bibinfo  {journal}
  {Rev. Mod. Phys.}\ }\textbf {\bibinfo {volume} {90}},\ \bibinfo {pages}
  {035005} (\bibinfo {year} {2018})}\BibitemShut {NoStop}%
\bibitem [{\citenamefont {Davis}\ \emph {et~al.}(2016)\citenamefont {Davis},
  \citenamefont {Bentsen},\ and\ \citenamefont {Schleier-Smith}}]{emily2016}%
  \BibitemOpen
  \bibfield  {author} {\bibinfo {author} {\bibfnamefont {Emily}\ \bibnamefont
  {Davis}}, \bibinfo {author} {\bibfnamefont {Gregory}\ \bibnamefont
  {Bentsen}}, \ and\ \bibinfo {author} {\bibfnamefont {Monika}\ \bibnamefont
  {Schleier-Smith}},\ }\bibfield  {title} {\enquote {\bibinfo {title}
  {Approaching the heisenberg limit without single-particle detection},}\
  }\href {\doibase 10.1103/PhysRevLett.116.053601} {\bibfield  {journal}
  {\bibinfo  {journal} {Phys. Rev. Lett.}\ }\textbf {\bibinfo {volume} {116}},\
  \bibinfo {pages} {053601} (\bibinfo {year} {2016})}\BibitemShut {NoStop}%
\bibitem [{\citenamefont {Fr\"owis}\ \emph {et~al.}(2016)\citenamefont
  {Fr\"owis}, \citenamefont {Sekatski},\ and\ \citenamefont
  {D\"ur}}]{frowis2016}%
  \BibitemOpen
  \bibfield  {author} {\bibinfo {author} {\bibfnamefont {Florian}\ \bibnamefont
  {Fr\"owis}}, \bibinfo {author} {\bibfnamefont {Pavel}\ \bibnamefont
  {Sekatski}}, \ and\ \bibinfo {author} {\bibfnamefont {Wolfgang}\ \bibnamefont
  {D\"ur}},\ }\bibfield  {title} {\enquote {\bibinfo {title} {Detecting large
  quantum fisher information with finite measurement precision},}\ }\href
  {\doibase 10.1103/PhysRevLett.116.090801} {\bibfield  {journal} {\bibinfo
  {journal} {Phys. Rev. Lett.}\ }\textbf {\bibinfo {volume} {116}},\ \bibinfo
  {pages} {090801} (\bibinfo {year} {2016})}\BibitemShut {NoStop}%
\bibitem [{\citenamefont {Macr\`{\i}}\ \emph {et~al.}(2016)\citenamefont
  {Macr\`{\i}}, \citenamefont {Smerzi},\ and\ \citenamefont
  {Pezz\`e}}]{macri2016}%
  \BibitemOpen
  \bibfield  {author} {\bibinfo {author} {\bibfnamefont {Tommaso}\ \bibnamefont
  {Macr\`{\i}}}, \bibinfo {author} {\bibfnamefont {Augusto}\ \bibnamefont
  {Smerzi}}, \ and\ \bibinfo {author} {\bibfnamefont {Luca}\ \bibnamefont
  {Pezz\`e}},\ }\bibfield  {title} {\enquote {\bibinfo {title} {Loschmidt echo
  for quantum metrology},}\ }\href {\doibase 10.1103/PhysRevA.94.010102}
  {\bibfield  {journal} {\bibinfo  {journal} {Phys. Rev. A}\ }\textbf {\bibinfo
  {volume} {94}},\ \bibinfo {pages} {010102(R)} (\bibinfo {year}
  {2016})}\BibitemShut {NoStop}%
\bibitem [{\citenamefont {Hosten}\ \emph {et~al.}(2016)\citenamefont {Hosten},
  \citenamefont {Krishnakumar}, \citenamefont {Engelsen},\ and\ \citenamefont
  {Kasevich}}]{hosten2016quantum}%
  \BibitemOpen
  \bibfield  {author} {\bibinfo {author} {\bibfnamefont {Onur}\ \bibnamefont
  {Hosten}}, \bibinfo {author} {\bibfnamefont {Radha}\ \bibnamefont
  {Krishnakumar}}, \bibinfo {author} {\bibfnamefont {Nils~J}\ \bibnamefont
  {Engelsen}}, \ and\ \bibinfo {author} {\bibfnamefont {Mark~A}\ \bibnamefont
  {Kasevich}},\ }\bibfield  {title} {\enquote {\bibinfo {title} {Quantum phase
  magnification},}\ }\href
  {https://www.science.org/doi/10.1126/science.aaf3397} {\bibfield  {journal}
  {\bibinfo  {journal} {Science}\ }\textbf {\bibinfo {volume} {352}},\ \bibinfo
  {pages} {1552--1555} (\bibinfo {year} {2016})}\BibitemShut {NoStop}%
\bibitem [{\citenamefont {Burd}\ \emph {et~al.}(2019)\citenamefont {Burd},
  \citenamefont {Srinivas}, \citenamefont {Bollinger}, \citenamefont {Wilson},
  \citenamefont {Wineland}, \citenamefont {Leibfried}, \citenamefont
  {Slichter},\ and\ \citenamefont {Allcock}}]{burd2019}%
  \BibitemOpen
  \bibfield  {author} {\bibinfo {author} {\bibfnamefont {S.~C.}\ \bibnamefont
  {Burd}}, \bibinfo {author} {\bibfnamefont {R.}~\bibnamefont {Srinivas}},
  \bibinfo {author} {\bibfnamefont {J.~J.}\ \bibnamefont {Bollinger}}, \bibinfo
  {author} {\bibfnamefont {A.~C.}\ \bibnamefont {Wilson}}, \bibinfo {author}
  {\bibfnamefont {D.~J.}\ \bibnamefont {Wineland}}, \bibinfo {author}
  {\bibfnamefont {D.}~\bibnamefont {Leibfried}}, \bibinfo {author}
  {\bibfnamefont {D.~H.}\ \bibnamefont {Slichter}}, \ and\ \bibinfo {author}
  {\bibfnamefont {D.~T.~C.}\ \bibnamefont {Allcock}},\ }\bibfield  {title}
  {\enquote {\bibinfo {title} {Quantum amplification of mechanical oscillator
  motion},}\ }\href {\doibase 10.1126/science.aaw2884} {\bibfield  {journal}
  {\bibinfo  {journal} {Science}\ }\textbf {\bibinfo {volume} {364}},\ \bibinfo
  {pages} {1163--1165} (\bibinfo {year} {2019})}\BibitemShut {NoStop}%
\bibitem [{\citenamefont {Gilmore}\ \emph {et~al.}(2021)\citenamefont
  {Gilmore}, \citenamefont {Affolter}, \citenamefont {Lewis-Swan},
  \citenamefont {Barberena}, \citenamefont {Jordan}, \citenamefont {Rey},\ and\
  \citenamefont {Bollinger}}]{gilmore2021}%
  \BibitemOpen
  \bibfield  {author} {\bibinfo {author} {\bibfnamefont {Kevin~A.}\
  \bibnamefont {Gilmore}}, \bibinfo {author} {\bibfnamefont {Matthew}\
  \bibnamefont {Affolter}}, \bibinfo {author} {\bibfnamefont {Robert~J.}\
  \bibnamefont {Lewis-Swan}}, \bibinfo {author} {\bibfnamefont {Diego}\
  \bibnamefont {Barberena}}, \bibinfo {author} {\bibfnamefont {Elena}\
  \bibnamefont {Jordan}}, \bibinfo {author} {\bibfnamefont {Ana~Maria}\
  \bibnamefont {Rey}}, \ and\ \bibinfo {author} {\bibfnamefont {John~J.}\
  \bibnamefont {Bollinger}},\ }\bibfield  {title} {\enquote {\bibinfo {title}
  {Quantum-enhanced sensing of displacements and electric fields with
  two-dimensional trapped-ion crystals},}\ }\href {\doibase
  10.1126/science.abi5226} {\bibfield  {journal} {\bibinfo  {journal}
  {Science}\ }\textbf {\bibinfo {volume} {373}},\ \bibinfo {pages} {673--678}
  (\bibinfo {year} {2021})}\BibitemShut {NoStop}%
\bibitem [{\citenamefont {Colombo}\ \emph {et~al.}(2022)\citenamefont
  {Colombo}, \citenamefont {Pedrozo-Pe{\~{n}}afiel}, \citenamefont
  {Adiyatullin}, \citenamefont {Li}, \citenamefont {Mendez}, \citenamefont
  {Shu},\ and\ \citenamefont {Vuleti{\'{c}}}}]{colombo2021}%
  \BibitemOpen
  \bibfield  {author} {\bibinfo {author} {\bibfnamefont {Simone}\ \bibnamefont
  {Colombo}}, \bibinfo {author} {\bibfnamefont {Edwin}\ \bibnamefont
  {Pedrozo-Pe{\~{n}}afiel}}, \bibinfo {author} {\bibfnamefont {Albert~F.}\
  \bibnamefont {Adiyatullin}}, \bibinfo {author} {\bibfnamefont {Zeyang}\
  \bibnamefont {Li}}, \bibinfo {author} {\bibfnamefont {Enrique}\ \bibnamefont
  {Mendez}}, \bibinfo {author} {\bibfnamefont {Chi}\ \bibnamefont {Shu}}, \
  and\ \bibinfo {author} {\bibfnamefont {Vladan}\ \bibnamefont
  {Vuleti{\'{c}}}},\ }\bibfield  {title} {\enquote {\bibinfo {title}
  {Time-reversal-based quantum metrology with many-body entangled states},}\
  }\href {\doibase 10.1038/s41567-022-01653-5} {\bibfield  {journal} {\bibinfo
  {journal} {Nat. Phys.}\ }\textbf {\bibinfo {volume} {18}},\ \bibinfo {pages}
  {925--930} (\bibinfo {year} {2022})}\BibitemShut {NoStop}%
\bibitem [{\citenamefont {Linnemann}\ \emph {et~al.}(2016)\citenamefont
  {Linnemann}, \citenamefont {Strobel}, \citenamefont {Muessel}, \citenamefont
  {Schulz}, \citenamefont {Lewis-Swan}, \citenamefont {Kheruntsyan},\ and\
  \citenamefont {Oberthaler}}]{linnemann2016}%
  \BibitemOpen
  \bibfield  {author} {\bibinfo {author} {\bibfnamefont {D.}~\bibnamefont
  {Linnemann}}, \bibinfo {author} {\bibfnamefont {H.}~\bibnamefont {Strobel}},
  \bibinfo {author} {\bibfnamefont {W.}~\bibnamefont {Muessel}}, \bibinfo
  {author} {\bibfnamefont {J.}~\bibnamefont {Schulz}}, \bibinfo {author}
  {\bibfnamefont {R.~J.}\ \bibnamefont {Lewis-Swan}}, \bibinfo {author}
  {\bibfnamefont {K.~V.}\ \bibnamefont {Kheruntsyan}}, \ and\ \bibinfo {author}
  {\bibfnamefont {M.~K.}\ \bibnamefont {Oberthaler}},\ }\bibfield  {title}
  {\enquote {\bibinfo {title} {Quantum-enhanced sensing based on time reversal
  of nonlinear dynamics},}\ }\href {\doibase 10.1103/PhysRevLett.117.013001}
  {\bibfield  {journal} {\bibinfo  {journal} {Phys. Rev. Lett.}\ }\textbf
  {\bibinfo {volume} {117}},\ \bibinfo {pages} {013001} (\bibinfo {year}
  {2016})}\BibitemShut {NoStop}%
\bibitem [{\citenamefont {Hudelist}\ \emph {et~al.}(2014)\citenamefont
  {Hudelist}, \citenamefont {Kong}, \citenamefont {Liu}, \citenamefont {Jing},
  \citenamefont {Ou},\ and\ \citenamefont {Zhang}}]{hudelist2014}%
  \BibitemOpen
  \bibfield  {author} {\bibinfo {author} {\bibfnamefont {F.}~\bibnamefont
  {Hudelist}}, \bibinfo {author} {\bibfnamefont {Jia}\ \bibnamefont {Kong}},
  \bibinfo {author} {\bibfnamefont {Cunjin}\ \bibnamefont {Liu}}, \bibinfo
  {author} {\bibfnamefont {Jietai}\ \bibnamefont {Jing}}, \bibinfo {author}
  {\bibfnamefont {Z.~Y.}\ \bibnamefont {Ou}}, \ and\ \bibinfo {author}
  {\bibfnamefont {Weiping}\ \bibnamefont {Zhang}},\ }\bibfield  {title}
  {\enquote {\bibinfo {title} {Quantum metrology with parametric
  amplifier-based photon correlation interferometers},}\ }\href {\doibase
  10.1038/ncomms4049} {\bibfield  {journal} {\bibinfo  {journal} {Nature
  Communications}\ }\textbf {\bibinfo {volume} {5}},\ \bibinfo {pages} {3049}
  (\bibinfo {year} {2014})}\BibitemShut {NoStop}%
\bibitem [{\citenamefont {Liu}\ \emph {et~al.}(2018)\citenamefont {Liu},
  \citenamefont {Li}, \citenamefont {Cui}, \citenamefont {Huo}, \citenamefont
  {Assad}, \citenamefont {Li},\ and\ \citenamefont {Ou}}]{Li2018}%
  \BibitemOpen
  \bibfield  {author} {\bibinfo {author} {\bibfnamefont {Yuhong}\ \bibnamefont
  {Liu}}, \bibinfo {author} {\bibfnamefont {Jiamin}\ \bibnamefont {Li}},
  \bibinfo {author} {\bibfnamefont {Liang}\ \bibnamefont {Cui}}, \bibinfo
  {author} {\bibfnamefont {Nan}\ \bibnamefont {Huo}}, \bibinfo {author}
  {\bibfnamefont {Syed~M.}\ \bibnamefont {Assad}}, \bibinfo {author}
  {\bibfnamefont {Xiaoying}\ \bibnamefont {Li}}, \ and\ \bibinfo {author}
  {\bibfnamefont {Z.~Y.}\ \bibnamefont {Ou}},\ }\bibfield  {title} {\enquote
  {\bibinfo {title} {Loss-tolerant quantum dense metrology with {SU}(1,1)
  interferometer},}\ }\href {\doibase 10.1364/OE.26.027705} {\bibfield
  {journal} {\bibinfo  {journal} {Opt. Express}\ }\textbf {\bibinfo {volume}
  {26}},\ \bibinfo {pages} {27705--27715} (\bibinfo {year} {2018})}\BibitemShut
  {NoStop}%
\bibitem [{\citenamefont {Sutton}\ and\ \citenamefont
  {Barto}(2018)}]{sutton2018}%
  \BibitemOpen
  \bibfield  {author} {\bibinfo {author} {\bibfnamefont {Richard~S}\
  \bibnamefont {Sutton}}\ and\ \bibinfo {author} {\bibfnamefont {Andrew~G}\
  \bibnamefont {Barto}},\ }\href@noop {} {\emph {\bibinfo {title}
  {Reinforcement learning: An introduction}}}\ (\bibinfo  {publisher} {MIT
  press},\ \bibinfo {year} {2018})\BibitemShut {NoStop}%
\bibitem [{\citenamefont {Bukov}\ \emph {et~al.}(2018)\citenamefont {Bukov},
  \citenamefont {Day}, \citenamefont {Sels}, \citenamefont {Weinberg},
  \citenamefont {Polkovnikov},\ and\ \citenamefont {Mehta}}]{bukov2018}%
  \BibitemOpen
  \bibfield  {author} {\bibinfo {author} {\bibfnamefont {Marin}\ \bibnamefont
  {Bukov}}, \bibinfo {author} {\bibfnamefont {Alexandre G.~R.}\ \bibnamefont
  {Day}}, \bibinfo {author} {\bibfnamefont {Dries}\ \bibnamefont {Sels}},
  \bibinfo {author} {\bibfnamefont {Phillip}\ \bibnamefont {Weinberg}},
  \bibinfo {author} {\bibfnamefont {Anatoli}\ \bibnamefont {Polkovnikov}}, \
  and\ \bibinfo {author} {\bibfnamefont {Pankaj}\ \bibnamefont {Mehta}},\
  }\bibfield  {title} {\enquote {\bibinfo {title} {Reinforcement learning in
  different phases of quantum control},}\ }\href {\doibase
  10.1103/PhysRevX.8.031086} {\bibfield  {journal} {\bibinfo  {journal} {Phys.
  Rev. X}\ }\textbf {\bibinfo {volume} {8}},\ \bibinfo {pages} {031086}
  (\bibinfo {year} {2018})}\BibitemShut {NoStop}%
\bibitem [{\citenamefont {Dalgaard}\ \emph {et~al.}(2020)\citenamefont
  {Dalgaard}, \citenamefont {Motzoi}, \citenamefont {S{\o}rensen},\ and\
  \citenamefont {Sherson}}]{dalgaard2020}%
  \BibitemOpen
  \bibfield  {author} {\bibinfo {author} {\bibfnamefont {Mogens}\ \bibnamefont
  {Dalgaard}}, \bibinfo {author} {\bibfnamefont {Felix}\ \bibnamefont
  {Motzoi}}, \bibinfo {author} {\bibfnamefont {Jens~Jakob}\ \bibnamefont
  {S{\o}rensen}}, \ and\ \bibinfo {author} {\bibfnamefont {Jacob}\ \bibnamefont
  {Sherson}},\ }\bibfield  {title} {\enquote {\bibinfo {title} {Global
  optimization of quantum dynamics with alphazero deep exploration},}\ }\href
  {http://dx.doi.org/10.1038/s41534-019-0241-0} {\bibfield  {journal} {\bibinfo
   {journal} {npj Quantum Information}\ }\textbf {\bibinfo {volume} {6}}
  (\bibinfo {year} {2020})}\BibitemShut {NoStop}%
\bibitem [{\citenamefont {Wauters}\ \emph {et~al.}(2020)\citenamefont
  {Wauters}, \citenamefont {Panizon}, \citenamefont {Mbeng},\ and\
  \citenamefont {Santoro}}]{wauters2020}%
  \BibitemOpen
  \bibfield  {author} {\bibinfo {author} {\bibfnamefont {Matteo~M.}\
  \bibnamefont {Wauters}}, \bibinfo {author} {\bibfnamefont {Emanuele}\
  \bibnamefont {Panizon}}, \bibinfo {author} {\bibfnamefont {Glen~B.}\
  \bibnamefont {Mbeng}}, \ and\ \bibinfo {author} {\bibfnamefont {Giuseppe~E.}\
  \bibnamefont {Santoro}},\ }\bibfield  {title} {\enquote {\bibinfo {title}
  {Reinforcement-learning-assisted quantum optimization},}\ }\href {\doibase
  10.1103/PhysRevResearch.2.033446} {\bibfield  {journal} {\bibinfo  {journal}
  {Phys. Rev. Research}\ }\textbf {\bibinfo {volume} {2}},\ \bibinfo {pages}
  {033446} (\bibinfo {year} {2020})}\BibitemShut {NoStop}%
\bibitem [{\citenamefont {Guo}\ \emph {et~al.}(2021)\citenamefont {Guo},
  \citenamefont {Chen}, \citenamefont {Liu}, \citenamefont {Xue}, \citenamefont
  {Chen}, \citenamefont {Cao}, \citenamefont {Mao}, \citenamefont {Tey},\ and\
  \citenamefont {You}}]{guo2021}%
  \BibitemOpen
  \bibfield  {author} {\bibinfo {author} {\bibfnamefont {Shuai-Feng}\
  \bibnamefont {Guo}}, \bibinfo {author} {\bibfnamefont {Feng}\ \bibnamefont
  {Chen}}, \bibinfo {author} {\bibfnamefont {Qi}~\bibnamefont {Liu}}, \bibinfo
  {author} {\bibfnamefont {Ming}\ \bibnamefont {Xue}}, \bibinfo {author}
  {\bibfnamefont {Jun-Jie}\ \bibnamefont {Chen}}, \bibinfo {author}
  {\bibfnamefont {Jia-Hao}\ \bibnamefont {Cao}}, \bibinfo {author}
  {\bibfnamefont {Tian-Wei}\ \bibnamefont {Mao}}, \bibinfo {author}
  {\bibfnamefont {Meng~Khoon}\ \bibnamefont {Tey}}, \ and\ \bibinfo {author}
  {\bibfnamefont {Li}~\bibnamefont {You}},\ }\bibfield  {title} {\enquote
  {\bibinfo {title} {Faster state preparation across quantum phase transition
  assisted by reinforcement learning},}\ }\href {\doibase
  10.1103/PhysRevLett.126.060401} {\bibfield  {journal} {\bibinfo  {journal}
  {Phys. Rev. Lett.}\ }\textbf {\bibinfo {volume} {126}},\ \bibinfo {pages}
  {060401} (\bibinfo {year} {2021})}\BibitemShut {NoStop}%
\bibitem [{\citenamefont {Yao}\ \emph {et~al.}(2021)\citenamefont {Yao},
  \citenamefont {Lin},\ and\ \citenamefont {Bukov}}]{yao2020a}%
  \BibitemOpen
  \bibfield  {author} {\bibinfo {author} {\bibfnamefont {Jiahao}\ \bibnamefont
  {Yao}}, \bibinfo {author} {\bibfnamefont {Lin}\ \bibnamefont {Lin}}, \ and\
  \bibinfo {author} {\bibfnamefont {Marin}\ \bibnamefont {Bukov}},\ }\bibfield
  {title} {\enquote {\bibinfo {title} {Reinforcement learning for many-body
  ground-state preparation inspired by counterdiabatic driving},}\ }\href
  {\doibase 10.1103/PhysRevX.11.031070} {\bibfield  {journal} {\bibinfo
  {journal} {Phys. Rev. X}\ }\textbf {\bibinfo {volume} {11}},\ \bibinfo
  {pages} {031070} (\bibinfo {year} {2021})}\BibitemShut {NoStop}%
\bibitem [{\citenamefont {Haug}\ \emph {et~al.}(2020)\citenamefont {Haug},
  \citenamefont {Mok}, \citenamefont {You}, \citenamefont {Zhang},
  \citenamefont {Png},\ and\ \citenamefont {Kwek}}]{Haug_2020}%
  \BibitemOpen
  \bibfield  {author} {\bibinfo {author} {\bibfnamefont {Tobias}\ \bibnamefont
  {Haug}}, \bibinfo {author} {\bibfnamefont {Wai-Keong}\ \bibnamefont {Mok}},
  \bibinfo {author} {\bibfnamefont {Jia-Bin}\ \bibnamefont {You}}, \bibinfo
  {author} {\bibfnamefont {Wenzu}\ \bibnamefont {Zhang}}, \bibinfo {author}
  {\bibfnamefont {Ching~Eng}\ \bibnamefont {Png}}, \ and\ \bibinfo {author}
  {\bibfnamefont {Leong-Chuan}\ \bibnamefont {Kwek}},\ }\bibfield  {title}
  {\enquote {\bibinfo {title} {Classifying global state preparation via deep
  reinforcement learning},}\ }\href {\doibase 10.1088/2632-2153/abc81f}
  {\bibfield  {journal} {\bibinfo  {journal} {Machine Learning: Science and
  Technology}\ }\textbf {\bibinfo {volume} {2}},\ \bibinfo {pages} {01LT02}
  (\bibinfo {year} {2020})}\BibitemShut {NoStop}%
\bibitem [{\citenamefont {Borah}\ \emph {et~al.}(2021)\citenamefont {Borah},
  \citenamefont {Sarma}, \citenamefont {Kewming}, \citenamefont {Milburn},\
  and\ \citenamefont {Twamley}}]{borah2021}%
  \BibitemOpen
  \bibfield  {author} {\bibinfo {author} {\bibfnamefont {Sangkha}\ \bibnamefont
  {Borah}}, \bibinfo {author} {\bibfnamefont {Bijita}\ \bibnamefont {Sarma}},
  \bibinfo {author} {\bibfnamefont {Michael}\ \bibnamefont {Kewming}}, \bibinfo
  {author} {\bibfnamefont {Gerard~J.}\ \bibnamefont {Milburn}}, \ and\ \bibinfo
  {author} {\bibfnamefont {Jason}\ \bibnamefont {Twamley}},\ }\bibfield
  {title} {\enquote {\bibinfo {title} {Measurement-based feedback quantum
  control with deep reinforcement learning for a double-well nonlinear
  potential},}\ }\href {\doibase 10.1103/PhysRevLett.127.190403} {\bibfield
  {journal} {\bibinfo  {journal} {Phys. Rev. Lett.}\ }\textbf {\bibinfo
  {volume} {127}},\ \bibinfo {pages} {190403} (\bibinfo {year}
  {2021})}\BibitemShut {NoStop}%
\bibitem [{\citenamefont {Porotti}\ \emph {et~al.}(2022)\citenamefont
  {Porotti}, \citenamefont {Essig}, \citenamefont {Huard},\ and\ \citenamefont
  {Marquardt}}]{porotti2022deep}%
  \BibitemOpen
  \bibfield  {author} {\bibinfo {author} {\bibfnamefont {Riccardo}\
  \bibnamefont {Porotti}}, \bibinfo {author} {\bibfnamefont {Antoine}\
  \bibnamefont {Essig}}, \bibinfo {author} {\bibfnamefont {Benjamin}\
  \bibnamefont {Huard}}, \ and\ \bibinfo {author} {\bibfnamefont {Florian}\
  \bibnamefont {Marquardt}},\ }\bibfield  {title} {\enquote {\bibinfo {title}
  {Deep reinforcement learning for quantum state preparation with weak
  nonlinear measurements},}\ }\href@noop {} {\bibfield  {journal} {\bibinfo
  {journal} {Quantum}\ }\textbf {\bibinfo {volume} {6}},\ \bibinfo {pages}
  {747} (\bibinfo {year} {2022})}\BibitemShut {NoStop}%
\bibitem [{\citenamefont {An}\ and\ \citenamefont {Zhou}(2019)}]{an2019}%
  \BibitemOpen
  \bibfield  {author} {\bibinfo {author} {\bibfnamefont {Zheng}\ \bibnamefont
  {An}}\ and\ \bibinfo {author} {\bibfnamefont {D.~L.}\ \bibnamefont {Zhou}},\
  }\bibfield  {title} {\enquote {\bibinfo {title} {Deep reinforcement learning
  for quantum gate control},}\ }\href {\doibase 10.1209/0295-5075/126/60002}
  {\bibfield  {journal} {\bibinfo  {journal} {{EPL} (Europhysics Letters)}\
  }\textbf {\bibinfo {volume} {126}},\ \bibinfo {pages} {60002} (\bibinfo
  {year} {2019})}\BibitemShut {NoStop}%
\bibitem [{\citenamefont {Niu}\ \emph {et~al.}(2019)\citenamefont {Niu},
  \citenamefont {Boixo}, \citenamefont {Smelyanskiy},\ and\ \citenamefont
  {Neven}}]{niu2019}%
  \BibitemOpen
  \bibfield  {author} {\bibinfo {author} {\bibfnamefont {Murphy~Yuezhen}\
  \bibnamefont {Niu}}, \bibinfo {author} {\bibfnamefont {Sergio}\ \bibnamefont
  {Boixo}}, \bibinfo {author} {\bibfnamefont {Vadim~N.}\ \bibnamefont
  {Smelyanskiy}}, \ and\ \bibinfo {author} {\bibfnamefont {Hartmut}\
  \bibnamefont {Neven}},\ }\bibfield  {title} {\enquote {\bibinfo {title}
  {Universal quantum control through deep reinforcement learning},}\ }\href
  {http://dx.doi.org/10.1038/s41534-019-0141-3} {\bibfield  {journal} {\bibinfo
   {journal} {npj Quantum Information}\ }\textbf {\bibinfo {volume} {5}}
  (\bibinfo {year} {2019})}\BibitemShut {NoStop}%
\bibitem [{\citenamefont {Wang}\ \emph {et~al.}(2020)\citenamefont {Wang},
  \citenamefont {Ashida},\ and\ \citenamefont {Ueda}}]{wang2020}%
  \BibitemOpen
  \bibfield  {author} {\bibinfo {author} {\bibfnamefont {Zhikang~T.}\
  \bibnamefont {Wang}}, \bibinfo {author} {\bibfnamefont {Yuto}\ \bibnamefont
  {Ashida}}, \ and\ \bibinfo {author} {\bibfnamefont {Masahito}\ \bibnamefont
  {Ueda}},\ }\bibfield  {title} {\enquote {\bibinfo {title} {Deep reinforcement
  learning control of quantum cartpoles},}\ }\href {\doibase
  10.1103/PhysRevLett.125.100401} {\bibfield  {journal} {\bibinfo  {journal}
  {Phys. Rev. Lett.}\ }\textbf {\bibinfo {volume} {125}},\ \bibinfo {pages}
  {100401} (\bibinfo {year} {2020})}\BibitemShut {NoStop}%
\bibitem [{\citenamefont {Lin}\ \emph {et~al.}(2020)\citenamefont {Lin},
  \citenamefont {Lai},\ and\ \citenamefont {Li}}]{PhysRevA.101.052327}%
  \BibitemOpen
  \bibfield  {author} {\bibinfo {author} {\bibfnamefont {Jian}\ \bibnamefont
  {Lin}}, \bibinfo {author} {\bibfnamefont {Zhong~Yuan}\ \bibnamefont {Lai}}, \
  and\ \bibinfo {author} {\bibfnamefont {Xiaopeng}\ \bibnamefont {Li}},\
  }\bibfield  {title} {\enquote {\bibinfo {title} {Quantum adiabatic algorithm
  design using reinforcement learning},}\ }\href {\doibase
  10.1103/PhysRevA.101.052327} {\bibfield  {journal} {\bibinfo  {journal}
  {Phys. Rev. A}\ }\textbf {\bibinfo {volume} {101}},\ \bibinfo {pages}
  {052327} (\bibinfo {year} {2020})}\BibitemShut {NoStop}%
\bibitem [{\citenamefont {Saggio}\ \emph {et~al.}(2021)\citenamefont {Saggio},
  \citenamefont {Asenbeck}, \citenamefont {Hamann}, \citenamefont
  {Str{\"o}mberg}, \citenamefont {Schiansky}, \citenamefont {Dunjko},
  \citenamefont {Friis}, \citenamefont {Harris}, \citenamefont {Hochberg},
  \citenamefont {Englund}, \citenamefont {W{\"o}lk}, \citenamefont {Briegel},\
  and\ \citenamefont {Walther}}]{Saggio2021}%
  \BibitemOpen
  \bibfield  {author} {\bibinfo {author} {\bibfnamefont {V.}~\bibnamefont
  {Saggio}}, \bibinfo {author} {\bibfnamefont {B.~E.}\ \bibnamefont
  {Asenbeck}}, \bibinfo {author} {\bibfnamefont {A.}~\bibnamefont {Hamann}},
  \bibinfo {author} {\bibfnamefont {T.}~\bibnamefont {Str{\"o}mberg}}, \bibinfo
  {author} {\bibfnamefont {P.}~\bibnamefont {Schiansky}}, \bibinfo {author}
  {\bibfnamefont {V.}~\bibnamefont {Dunjko}}, \bibinfo {author} {\bibfnamefont
  {N.}~\bibnamefont {Friis}}, \bibinfo {author} {\bibfnamefont {N.~C.}\
  \bibnamefont {Harris}}, \bibinfo {author} {\bibfnamefont {M.}~\bibnamefont
  {Hochberg}}, \bibinfo {author} {\bibfnamefont {D.}~\bibnamefont {Englund}},
  \bibinfo {author} {\bibfnamefont {S.}~\bibnamefont {W{\"o}lk}}, \bibinfo
  {author} {\bibfnamefont {H.~J.}\ \bibnamefont {Briegel}}, \ and\ \bibinfo
  {author} {\bibfnamefont {P.}~\bibnamefont {Walther}},\ }\bibfield  {title}
  {\enquote {\bibinfo {title} {Experimental quantum speed-up in
  reinforcement learning agents},}\ }\href {\doibase
  10.1038/s41586-021-03242-7} {\bibfield  {journal} {\bibinfo  {journal}
  {Nature}\ }\textbf {\bibinfo {volume} {591}},\ \bibinfo {pages} {229--233}
  (\bibinfo {year} {2021})}\BibitemShut {NoStop}%
\bibitem [{\citenamefont {Xu}\ \emph {et~al.}(2019)\citenamefont {Xu},
  \citenamefont {Li}, \citenamefont {Liu}, \citenamefont {Wang}, \citenamefont
  {Yuan},\ and\ \citenamefont {Wang}}]{Xu2019}%
  \BibitemOpen
  \bibfield  {author} {\bibinfo {author} {\bibfnamefont {Han}\ \bibnamefont
  {Xu}}, \bibinfo {author} {\bibfnamefont {Junning}\ \bibnamefont {Li}},
  \bibinfo {author} {\bibfnamefont {Liqiang}\ \bibnamefont {Liu}}, \bibinfo
  {author} {\bibfnamefont {Yu}~\bibnamefont {Wang}}, \bibinfo {author}
  {\bibfnamefont {Haidong}\ \bibnamefont {Yuan}}, \ and\ \bibinfo {author}
  {\bibfnamefont {Xin}\ \bibnamefont {Wang}},\ }\bibfield  {title} {\enquote
  {\bibinfo {title} {Generalizable control for quantum parameter estimation
  through reinforcement learning},}\ }\href {\doibase
  10.1038/s41534-019-0198-z} {\bibfield  {journal} {\bibinfo  {journal} {npj
  Quantum Information}\ }\textbf {\bibinfo {volume} {5}},\ \bibinfo {pages}
  {82} (\bibinfo {year} {2019})}\BibitemShut {NoStop}%
\bibitem [{\citenamefont {Schuff}\ \emph {et~al.}(2020)\citenamefont {Schuff},
  \citenamefont {Fiderer},\ and\ \citenamefont {Braun}}]{Schuff_2020}%
  \BibitemOpen
  \bibfield  {author} {\bibinfo {author} {\bibfnamefont {Jonas}\ \bibnamefont
  {Schuff}}, \bibinfo {author} {\bibfnamefont {Lukas~J}\ \bibnamefont
  {Fiderer}}, \ and\ \bibinfo {author} {\bibfnamefont {Daniel}\ \bibnamefont
  {Braun}},\ }\bibfield  {title} {\enquote {\bibinfo {title} {Improving the
  dynamics of quantum sensors with reinforcement learning},}\ }\href {\doibase
  10.1088/1367-2630/ab6f1f} {\bibfield  {journal} {\bibinfo  {journal} {New
  Journal of Physics}\ }\textbf {\bibinfo {volume} {22}},\ \bibinfo {pages}
  {035001} (\bibinfo {year} {2020})}\BibitemShut {NoStop}%
\bibitem [{\citenamefont {Chih}\ and\ \citenamefont
  {Holland}(2021)}]{PhysRevResearch.3.033279}%
  \BibitemOpen
  \bibfield  {author} {\bibinfo {author} {\bibfnamefont {Liang-Ying}\
  \bibnamefont {Chih}}\ and\ \bibinfo {author} {\bibfnamefont {Murray}\
  \bibnamefont {Holland}},\ }\bibfield  {title} {\enquote {\bibinfo {title}
  {Reinforcement-learning-based matter-wave interferometer in a shaken optical
  lattice},}\ }\href {\doibase 10.1103/PhysRevResearch.3.033279} {\bibfield
  {journal} {\bibinfo  {journal} {Phys. Rev. Research}\ }\textbf {\bibinfo
  {volume} {3}},\ \bibinfo {pages} {033279} (\bibinfo {year}
  {2021})}\BibitemShut {NoStop}%
\bibitem [{\citenamefont {Qiu}\ \emph {et~al.}(2022)\citenamefont {Qiu},
  \citenamefont {Zhuang}, \citenamefont {Huang},\ and\ \citenamefont
  {Lee}}]{2022arXiv220300189Q}%
  \BibitemOpen
  \bibfield  {author} {\bibinfo {author} {\bibfnamefont {Yuxiang}\ \bibnamefont
  {Qiu}}, \bibinfo {author} {\bibfnamefont {Min}\ \bibnamefont {Zhuang}},
  \bibinfo {author} {\bibfnamefont {Jiahao}\ \bibnamefont {Huang}}, \ and\
  \bibinfo {author} {\bibfnamefont {Chaohong}\ \bibnamefont {Lee}},\ }\bibfield
   {title} {\enquote {\bibinfo {title} {Efficient and robust entanglement
  generation with deep reinforcement learning for quantum metrology},}\ }\href
  {\doibase 10.1088/1367-2630/ac8285} {\bibfield  {journal} {\bibinfo
  {journal} {New Journal of Physics}\ }\textbf {\bibinfo {volume} {24}},\
  \bibinfo {pages} {083011} (\bibinfo {year} {2022})}\BibitemShut {NoStop}%
\bibitem [{\citenamefont {F\"osel}\ \emph {et~al.}(2018)\citenamefont
  {F\"osel}, \citenamefont {Tighineanu}, \citenamefont {Weiss},\ and\
  \citenamefont {Marquardt}}]{fosel2018}%
  \BibitemOpen
  \bibfield  {author} {\bibinfo {author} {\bibfnamefont {Thomas}\ \bibnamefont
  {F\"osel}}, \bibinfo {author} {\bibfnamefont {Petru}\ \bibnamefont
  {Tighineanu}}, \bibinfo {author} {\bibfnamefont {Talitha}\ \bibnamefont
  {Weiss}}, \ and\ \bibinfo {author} {\bibfnamefont {Florian}\ \bibnamefont
  {Marquardt}},\ }\bibfield  {title} {\enquote {\bibinfo {title} {Reinforcement
  learning with neural networks for quantum feedback},}\ }\href {\doibase
  10.1103/PhysRevX.8.031084} {\bibfield  {journal} {\bibinfo  {journal} {Phys.
  Rev. X}\ }\textbf {\bibinfo {volume} {8}},\ \bibinfo {pages} {031084}
  (\bibinfo {year} {2018})}\BibitemShut {NoStop}%
\bibitem [{\citenamefont {Nautrup}\ \emph {et~al.}(2019)\citenamefont
  {Nautrup}, \citenamefont {Delfosse}, \citenamefont {Dunjko}, \citenamefont
  {Briegel},\ and\ \citenamefont {Friis}}]{nautrup2019}%
  \BibitemOpen
  \bibfield  {author} {\bibinfo {author} {\bibfnamefont {Hendrik~Poulsen}\
  \bibnamefont {Nautrup}}, \bibinfo {author} {\bibfnamefont {Nicolas}\
  \bibnamefont {Delfosse}}, \bibinfo {author} {\bibfnamefont {Vedran}\
  \bibnamefont {Dunjko}}, \bibinfo {author} {\bibfnamefont {Hans~J.}\
  \bibnamefont {Briegel}}, \ and\ \bibinfo {author} {\bibfnamefont {Nicolai}\
  \bibnamefont {Friis}},\ }\bibfield  {title} {\enquote {\bibinfo {title}
  {Optimizing {Q}uantum {E}rror {C}orrection {C}odes with {R}einforcement
  {L}earning},}\ }\href {\doibase 10.22331/q-2019-12-16-215} {\bibfield
  {journal} {\bibinfo  {journal} {{Quantum}}\ }\textbf {\bibinfo {volume}
  {3}},\ \bibinfo {pages} {215} (\bibinfo {year} {2019})}\BibitemShut {NoStop}%
\bibitem [{\citenamefont {Andreasson}\ \emph {et~al.}(2019)\citenamefont
  {Andreasson}, \citenamefont {Johansson}, \citenamefont {Liljestrand},\ and\
  \citenamefont {Granath}}]{Andreasson2019quantumerror}%
  \BibitemOpen
  \bibfield  {author} {\bibinfo {author} {\bibfnamefont {Philip}\ \bibnamefont
  {Andreasson}}, \bibinfo {author} {\bibfnamefont {Joel}\ \bibnamefont
  {Johansson}}, \bibinfo {author} {\bibfnamefont {Simon}\ \bibnamefont
  {Liljestrand}}, \ and\ \bibinfo {author} {\bibfnamefont {Mats}\ \bibnamefont
  {Granath}},\ }\bibfield  {title} {\enquote {\bibinfo {title} {Quantum error
  correction for the toric code using deep reinforcement learning},}\ }\href
  {\doibase 10.22331/q-2019-09-02-183} {\bibfield  {journal} {\bibinfo
  {journal} {{Quantum}}\ }\textbf {\bibinfo {volume} {3}},\ \bibinfo {pages}
  {183} (\bibinfo {year} {2019})}\BibitemShut {NoStop}%
\bibitem [{\citenamefont {Zhang}\ \emph {et~al.}(2020)\citenamefont {Zhang},
  \citenamefont {Zheng}, \citenamefont {Zhang},\ and\ \citenamefont
  {Deng}}]{PhysRevLett.125.170501}%
  \BibitemOpen
  \bibfield  {author} {\bibinfo {author} {\bibfnamefont {Yuan-Hang}\
  \bibnamefont {Zhang}}, \bibinfo {author} {\bibfnamefont {Pei-Lin}\
  \bibnamefont {Zheng}}, \bibinfo {author} {\bibfnamefont {Yi}~\bibnamefont
  {Zhang}}, \ and\ \bibinfo {author} {\bibfnamefont {Dong-Ling}\ \bibnamefont
  {Deng}},\ }\bibfield  {title} {\enquote {\bibinfo {title} {Topological
  quantum compiling with reinforcement learning},}\ }\href {\doibase
  10.1103/PhysRevLett.125.170501} {\bibfield  {journal} {\bibinfo  {journal}
  {Phys. Rev. Lett.}\ }\textbf {\bibinfo {volume} {125}},\ \bibinfo {pages}
  {170501} (\bibinfo {year} {2020})}\BibitemShut {NoStop}%
\bibitem [{\citenamefont {Moro}\ \emph {et~al.}(2021)\citenamefont {Moro},
  \citenamefont {Paris}, \citenamefont {Restelli},\ and\ \citenamefont
  {Prati}}]{Moro2021}%
  \BibitemOpen
  \bibfield  {author} {\bibinfo {author} {\bibfnamefont {Lorenzo}\ \bibnamefont
  {Moro}}, \bibinfo {author} {\bibfnamefont {Matteo G.~A.}\ \bibnamefont
  {Paris}}, \bibinfo {author} {\bibfnamefont {Marcello}\ \bibnamefont
  {Restelli}}, \ and\ \bibinfo {author} {\bibfnamefont {Enrico}\ \bibnamefont
  {Prati}},\ }\bibfield  {title} {\enquote {\bibinfo {title} {Quantum compiling
  by deep reinforcement learning},}\ }\href {\doibase
  10.1038/s42005-021-00684-3} {\bibfield  {journal} {\bibinfo  {journal}
  {Communications Physics}\ }\textbf {\bibinfo {volume} {4}},\ \bibinfo {pages}
  {178} (\bibinfo {year} {2021})}\BibitemShut {NoStop}%
\bibitem [{\citenamefont {He}\ \emph {et~al.}(2021)\citenamefont {He},
  \citenamefont {Li}, \citenamefont {Zheng}, \citenamefont {Li},\ and\
  \citenamefont {Situ}}]{He_2021}%
  \BibitemOpen
  \bibfield  {author} {\bibinfo {author} {\bibfnamefont {Zhimin}\ \bibnamefont
  {He}}, \bibinfo {author} {\bibfnamefont {Lvzhou}\ \bibnamefont {Li}},
  \bibinfo {author} {\bibfnamefont {Shenggen}\ \bibnamefont {Zheng}}, \bibinfo
  {author} {\bibfnamefont {Yongyao}\ \bibnamefont {Li}}, \ and\ \bibinfo
  {author} {\bibfnamefont {Haozhen}\ \bibnamefont {Situ}},\ }\bibfield  {title}
  {\enquote {\bibinfo {title} {Variational quantum compiling with double
  q-learning},}\ }\href {\doibase 10.1088/1367-2630/abe0ae} {\bibfield
  {journal} {\bibinfo  {journal} {New Journal of Physics}\ }\textbf {\bibinfo
  {volume} {23}},\ \bibinfo {pages} {033002} (\bibinfo {year}
  {2021})}\BibitemShut {NoStop}%
\bibitem [{\citenamefont {Hoang}\ \emph {et~al.}(2013)\citenamefont {Hoang},
  \citenamefont {Gerving}, \citenamefont {Land}, \citenamefont {Anquez},
  \citenamefont {Hamley},\ and\ \citenamefont
  {Chapman}}]{PhysRevLett.111.090403}%
  \BibitemOpen
  \bibfield  {author} {\bibinfo {author} {\bibfnamefont {T.~M.}\ \bibnamefont
  {Hoang}}, \bibinfo {author} {\bibfnamefont {C.~S.}\ \bibnamefont {Gerving}},
  \bibinfo {author} {\bibfnamefont {B.~J.}\ \bibnamefont {Land}}, \bibinfo
  {author} {\bibfnamefont {M.}~\bibnamefont {Anquez}}, \bibinfo {author}
  {\bibfnamefont {C.~D.}\ \bibnamefont {Hamley}}, \ and\ \bibinfo {author}
  {\bibfnamefont {M.~S.}\ \bibnamefont {Chapman}},\ }\bibfield  {title}
  {\enquote {\bibinfo {title} {Dynamic stabilization of a quantum many-body
  spin system},}\ }\href {\doibase 10.1103/PhysRevLett.111.090403} {\bibfield
  {journal} {\bibinfo  {journal} {Phys. Rev. Lett.}\ }\textbf {\bibinfo
  {volume} {111}},\ \bibinfo {pages} {090403} (\bibinfo {year}
  {2013})}\BibitemShut {NoStop}%
\bibitem [{\citenamefont {Law}\ \emph {et~al.}(1998)\citenamefont {Law},
  \citenamefont {Pu},\ and\ \citenamefont {Bigelow}}]{law1998}%
  \BibitemOpen
  \bibfield  {author} {\bibinfo {author} {\bibfnamefont {C.~K.}\ \bibnamefont
  {Law}}, \bibinfo {author} {\bibfnamefont {H.}~\bibnamefont {Pu}}, \ and\
  \bibinfo {author} {\bibfnamefont {N.~P.}\ \bibnamefont {Bigelow}},\
  }\bibfield  {title} {\enquote {\bibinfo {title} {Quantum {S}pins {M}ixing in
  {S}pinor {B}ose-{E}instein {C}ondensates},}\ }\href {\doibase
  10.1103/PhysRevLett.81.5257} {\bibfield  {journal} {\bibinfo  {journal}
  {Phys. Rev. Lett.}\ }\textbf {\bibinfo {volume} {81}},\ \bibinfo {pages}
  {5257--5261} (\bibinfo {year} {1998})}\BibitemShut {NoStop}%
\bibitem [{\citenamefont {Gerbier}\ \emph {et~al.}(2006)\citenamefont
  {Gerbier}, \citenamefont {Widera}, \citenamefont {F\"olling}, \citenamefont
  {Mandel},\ and\ \citenamefont {Bloch}}]{gerbier2006}%
  \BibitemOpen
  \bibfield  {author} {\bibinfo {author} {\bibfnamefont {Fabrice}\ \bibnamefont
  {Gerbier}}, \bibinfo {author} {\bibfnamefont {Artur}\ \bibnamefont {Widera}},
  \bibinfo {author} {\bibfnamefont {Simon}\ \bibnamefont {F\"olling}}, \bibinfo
  {author} {\bibfnamefont {Olaf}\ \bibnamefont {Mandel}}, \ and\ \bibinfo
  {author} {\bibfnamefont {Immanuel}\ \bibnamefont {Bloch}},\ }\bibfield
  {title} {\enquote {\bibinfo {title} {Resonant control of spin dynamics in
  ultracold quantum gases by microwave dressing},}\ }\href {\doibase
  10.1103/PhysRevA.73.041602} {\bibfield  {journal} {\bibinfo  {journal} {Phys.
  Rev. A}\ }\textbf {\bibinfo {volume} {73}},\ \bibinfo {pages} {041602(R)}
  (\bibinfo {year} {2006})}\BibitemShut {NoStop}%
\bibitem [{sup()}]{supp}%
  \BibitemOpen
  \href@noop {} {}\bibinfo {note} {See Supplemental Material which includes
  Refs.~\cite{guo2021,steel1998,sinatra2002, hamley2012,
  liu2021,ppo2017,he2015,ppo,gym,luo2017,
  zou2018,hoang2016,arlt2010,holland1993,JOHANSSON20131234} for (i) a detailed
  description of reinforcement learning task in Sec. I and transfer learning in
  Sec. II; (ii) simulation of dissipative systems in Sec. III; (iii)
  experimental methods in Sec. IV; (iv) more data in Sec. V; (v) a brief
  introduction to spin-1 Dicke state in Sec. VI; (vi) a comparison of our
  protocol with time-reversal protocol in Sec. VII, and (vii) a study on
  nonlinear readout in a spin-1/2 system in Sec. VIII, which also includes a
  comparison of RL with traditional optimization methods.}\BibitemShut {Stop}%
\bibitem [{\citenamefont {Zou}\ \emph {et~al.}(2018)\citenamefont {Zou},
  \citenamefont {Wu}, \citenamefont {Liu}, \citenamefont {Luo}, \citenamefont
  {Guo}, \citenamefont {Cao}, \citenamefont {Tey},\ and\ \citenamefont
  {You}}]{zou2018}%
  \BibitemOpen
  \bibfield  {author} {\bibinfo {author} {\bibfnamefont {Yi-Quan}\ \bibnamefont
  {Zou}}, \bibinfo {author} {\bibfnamefont {Ling-Na}\ \bibnamefont {Wu}},
  \bibinfo {author} {\bibfnamefont {Qi}~\bibnamefont {Liu}}, \bibinfo {author}
  {\bibfnamefont {Xin-Yu}\ \bibnamefont {Luo}}, \bibinfo {author}
  {\bibfnamefont {Shuai-Feng}\ \bibnamefont {Guo}}, \bibinfo {author}
  {\bibfnamefont {Jia-Hao}\ \bibnamefont {Cao}}, \bibinfo {author}
  {\bibfnamefont {Meng~Khoon}\ \bibnamefont {Tey}}, \ and\ \bibinfo {author}
  {\bibfnamefont {Li}~\bibnamefont {You}},\ }\bibfield  {title} {\enquote
  {\bibinfo {title} {Beating the classical precision limit with spin-1 dicke
  states of more than 10,000 atoms},}\ }\href {\doibase
  10.1073/pnas.1715105115} {\bibfield  {journal} {\bibinfo  {journal} {Proc.
  Natl. Acad. Sci. U.S.A.}\ }\textbf {\bibinfo {volume} {115}},\ \bibinfo
  {pages} {6381--6385} (\bibinfo {year} {2018})}\BibitemShut {NoStop}%
\bibitem [{\citenamefont {Liu}\ \emph {et~al.}(2022)\citenamefont {Liu},
  \citenamefont {Wu}, \citenamefont {Cao}, \citenamefont {Mao}, \citenamefont
  {Li}, \citenamefont {Guo}, \citenamefont {Tey},\ and\ \citenamefont
  {You}}]{liu2021}%
  \BibitemOpen
  \bibfield  {author} {\bibinfo {author} {\bibfnamefont {Qi}~\bibnamefont
  {Liu}}, \bibinfo {author} {\bibfnamefont {Ling-Na}\ \bibnamefont {Wu}},
  \bibinfo {author} {\bibfnamefont {Jia-Hao}\ \bibnamefont {Cao}}, \bibinfo
  {author} {\bibfnamefont {Tian-Wei}\ \bibnamefont {Mao}}, \bibinfo {author}
  {\bibfnamefont {Xin-Wei}\ \bibnamefont {Li}}, \bibinfo {author}
  {\bibfnamefont {Shuai-Feng}\ \bibnamefont {Guo}}, \bibinfo {author}
  {\bibfnamefont {Meng~Khoon}\ \bibnamefont {Tey}}, \ and\ \bibinfo {author}
  {\bibfnamefont {Li}~\bibnamefont {You}},\ }\bibfield  {title} {\enquote
  {\bibinfo {title} {Nonlinear interferometry beyond classical limit enabled by
  cyclic dynamics},}\ }\href {https://doi.org/10.1038/s41567-021-01441-7}
  {\bibfield  {journal} {\bibinfo  {journal} {Nat. Phys.}\ }\textbf {\bibinfo
  {volume} {18}},\ \bibinfo {pages} {167--171} (\bibinfo {year}
  {2022})}\BibitemShut {NoStop}%
\bibitem [{\citenamefont {Steel}\ \emph {et~al.}(1998)\citenamefont {Steel},
  \citenamefont {Olsen}, \citenamefont {Plimak}, \citenamefont {Drummond},
  \citenamefont {Tan}, \citenamefont {Collett}, \citenamefont {Walls},\ and\
  \citenamefont {Graham}}]{steel1998}%
  \BibitemOpen
  \bibfield  {author} {\bibinfo {author} {\bibfnamefont {M.~J.}\ \bibnamefont
  {Steel}}, \bibinfo {author} {\bibfnamefont {M.~K.}\ \bibnamefont {Olsen}},
  \bibinfo {author} {\bibfnamefont {L.~I.}\ \bibnamefont {Plimak}}, \bibinfo
  {author} {\bibfnamefont {P.~D.}\ \bibnamefont {Drummond}}, \bibinfo {author}
  {\bibfnamefont {S.~M.}\ \bibnamefont {Tan}}, \bibinfo {author} {\bibfnamefont
  {M.~J.}\ \bibnamefont {Collett}}, \bibinfo {author} {\bibfnamefont {D.~F.}\
  \bibnamefont {Walls}}, \ and\ \bibinfo {author} {\bibfnamefont
  {R.}~\bibnamefont {Graham}},\ }\bibfield  {title} {\enquote {\bibinfo {title}
  {Dynamical quantum noise in trapped {B}ose-{E}instein condensates},}\ }\href
  {\doibase 10.1103/PhysRevA.58.4824} {\bibfield  {journal} {\bibinfo
  {journal} {Phys. Rev. A}\ }\textbf {\bibinfo {volume} {58}},\ \bibinfo
  {pages} {4824--4835} (\bibinfo {year} {1998})}\BibitemShut {NoStop}%
\bibitem [{\citenamefont {Sinatra}\ \emph {et~al.}(2002)\citenamefont
  {Sinatra}, \citenamefont {Lobo},\ and\ \citenamefont {Castin}}]{sinatra2002}%
  \BibitemOpen
  \bibfield  {author} {\bibinfo {author} {\bibfnamefont {Alice}\ \bibnamefont
  {Sinatra}}, \bibinfo {author} {\bibfnamefont {Carlos}\ \bibnamefont {Lobo}},
  \ and\ \bibinfo {author} {\bibfnamefont {Yvan}\ \bibnamefont {Castin}},\
  }\bibfield  {title} {\enquote {\bibinfo {title} {The truncated wigner method
  for {B}ose-condensed gases: limits of validity and applications},}\ }\href
  {\doibase 10.1088/0953-4075/35/17/301} {\bibfield  {journal} {\bibinfo
  {journal} {J. Phys. B: At. Mol. Opt.}\ }\textbf {\bibinfo {volume} {35}},\
  \bibinfo {pages} {3599--3631} (\bibinfo {year} {2002})}\BibitemShut {NoStop}%
\bibitem [{\citenamefont {Norrie}\ \emph {et~al.}(2006)\citenamefont {Norrie},
  \citenamefont {Ballagh},\ and\ \citenamefont {Gardiner}}]{norrie2006}%
  \BibitemOpen
  \bibfield  {author} {\bibinfo {author} {\bibfnamefont {A.~A.}\ \bibnamefont
  {Norrie}}, \bibinfo {author} {\bibfnamefont {R.~J.}\ \bibnamefont {Ballagh}},
  \ and\ \bibinfo {author} {\bibfnamefont {C.~W.}\ \bibnamefont {Gardiner}},\
  }\bibfield  {title} {\enquote {\bibinfo {title} {Quantum turbulence and
  correlations in bose-einstein condensate collisions},}\ }\href {\doibase
  10.1103/PhysRevA.73.043617} {\bibfield  {journal} {\bibinfo  {journal} {Phys.
  Rev. A}\ }\textbf {\bibinfo {volume} {73}},\ \bibinfo {pages} {043617}
  (\bibinfo {year} {2006})}\BibitemShut {NoStop}%
\bibitem [{\citenamefont {Opanchuk}\ \emph {et~al.}(2012)\citenamefont
  {Opanchuk}, \citenamefont {Egorov}, \citenamefont {Hoffmann}, \citenamefont
  {Sidorov},\ and\ \citenamefont {Drummond}}]{opanchuk2012}%
  \BibitemOpen
  \bibfield  {author} {\bibinfo {author} {\bibfnamefont {B.}~\bibnamefont
  {Opanchuk}}, \bibinfo {author} {\bibfnamefont {M.}~\bibnamefont {Egorov}},
  \bibinfo {author} {\bibfnamefont {S.}~\bibnamefont {Hoffmann}}, \bibinfo
  {author} {\bibfnamefont {A.~I.}\ \bibnamefont {Sidorov}}, \ and\ \bibinfo
  {author} {\bibfnamefont {P.~D.}\ \bibnamefont {Drummond}},\ }\bibfield
  {title} {\enquote {\bibinfo {title} {Quantum noise in three-dimensional bec
  interferometry},}\ }\href {\doibase 10.1209/0295-5075/97/50003} {\bibfield
  {journal} {\bibinfo  {journal} {EPL (Europhysics Letters)}\ }\textbf
  {\bibinfo {volume} {97}},\ \bibinfo {pages} {50003} (\bibinfo {year}
  {2012})}\BibitemShut {NoStop}%
\bibitem [{\citenamefont {Drummond}\ and\ \citenamefont
  {Opanchuk}(2017)}]{drummond2017}%
  \BibitemOpen
  \bibfield  {author} {\bibinfo {author} {\bibfnamefont {Peter~D.}\
  \bibnamefont {Drummond}}\ and\ \bibinfo {author} {\bibfnamefont {Bogdan}\
  \bibnamefont {Opanchuk}},\ }\bibfield  {title} {\enquote {\bibinfo {title}
  {Truncated wigner dynamics and conservation laws},}\ }\href {\doibase
  10.1103/PhysRevA.96.043616} {\bibfield  {journal} {\bibinfo  {journal} {Phys.
  Rev. A}\ }\textbf {\bibinfo {volume} {96}},\ \bibinfo {pages} {043616}
  (\bibinfo {year} {2017})}\BibitemShut {NoStop}%
\bibitem [{\citenamefont {Johnson}\ \emph {et~al.}(2017)\citenamefont
  {Johnson}, \citenamefont {Szigeti}, \citenamefont {Schemmer},\ and\
  \citenamefont {Bouchoule}}]{johnson2017}%
  \BibitemOpen
  \bibfield  {author} {\bibinfo {author} {\bibfnamefont {A.}~\bibnamefont
  {Johnson}}, \bibinfo {author} {\bibfnamefont {S.~S.}\ \bibnamefont
  {Szigeti}}, \bibinfo {author} {\bibfnamefont {M.}~\bibnamefont {Schemmer}}, \
  and\ \bibinfo {author} {\bibfnamefont {I.}~\bibnamefont {Bouchoule}},\
  }\bibfield  {title} {\enquote {\bibinfo {title} {Long-lived nonthermal states
  realized by atom losses in one-dimensional quasicondensates},}\ }\href
  {\doibase 10.1103/PhysRevA.96.013623} {\bibfield  {journal} {\bibinfo
  {journal} {Phys. Rev. A}\ }\textbf {\bibinfo {volume} {96}},\ \bibinfo
  {pages} {013623} (\bibinfo {year} {2017})}\BibitemShut {NoStop}%
\bibitem [{\citenamefont {Gerving}\ \emph {et~al.}(2012)\citenamefont
  {Gerving}, \citenamefont {Hoang}, \citenamefont {Land}, \citenamefont
  {Anquez}, \citenamefont {Hamley},\ and\ \citenamefont
  {Chapman}}]{gerving2012}%
  \BibitemOpen
  \bibfield  {author} {\bibinfo {author} {\bibfnamefont {C.S.}\ \bibnamefont
  {Gerving}}, \bibinfo {author} {\bibfnamefont {T.M.}\ \bibnamefont {Hoang}},
  \bibinfo {author} {\bibfnamefont {B.J.}\ \bibnamefont {Land}}, \bibinfo
  {author} {\bibfnamefont {M.}~\bibnamefont {Anquez}}, \bibinfo {author}
  {\bibfnamefont {C.D.}\ \bibnamefont {Hamley}}, \ and\ \bibinfo {author}
  {\bibfnamefont {M.S.}\ \bibnamefont {Chapman}},\ }\bibfield  {title}
  {\enquote {\bibinfo {title} {Non-equilibrium dynamics of an unstable quantum
  pendulum explored in a spin-1 bose--einstein condensate},}\ }\href
  {http://dx.doi.org/10.1038/ncomms2179} {\bibfield  {journal} {\bibinfo
  {journal} {Nature Communications}\ }\textbf {\bibinfo {volume} {3}},\
  \bibinfo {pages} {1169} (\bibinfo {year} {2012})}\BibitemShut {NoStop}%
\bibitem [{\citenamefont {Hamley}\ \emph {et~al.}(2012)\citenamefont {Hamley},
  \citenamefont {Gerving}, \citenamefont {Hoang}, \citenamefont {Bookjans},\
  and\ \citenamefont {Chapman}}]{hamley2012}%
  \BibitemOpen
  \bibfield  {author} {\bibinfo {author} {\bibfnamefont {C.~D.}\ \bibnamefont
  {Hamley}}, \bibinfo {author} {\bibfnamefont {C.~S.}\ \bibnamefont {Gerving}},
  \bibinfo {author} {\bibfnamefont {T.~M.}\ \bibnamefont {Hoang}}, \bibinfo
  {author} {\bibfnamefont {E.~M.}\ \bibnamefont {Bookjans}}, \ and\ \bibinfo
  {author} {\bibfnamefont {M.~S.}\ \bibnamefont {Chapman}},\ }\bibfield
  {title} {\enquote {\bibinfo {title} {Spin-nematic squeezed vacuum in a
  quantum gas},}\ }\href {\doibase 10.1038/nphys2245} {\bibfield  {journal}
  {\bibinfo  {journal} {Nat. Phys.}\ }\textbf {\bibinfo {volume} {8}},\
  \bibinfo {pages} {305--308} (\bibinfo {year} {2012})}\BibitemShut {NoStop}%
\bibitem [{\citenamefont {Zhao}\ \emph {et~al.}(2014)\citenamefont {Zhao},
  \citenamefont {Jiang}, \citenamefont {Tang}, \citenamefont {Webb},\ and\
  \citenamefont {Liu}}]{zhao2014}%
  \BibitemOpen
  \bibfield  {author} {\bibinfo {author} {\bibfnamefont {L.}~\bibnamefont
  {Zhao}}, \bibinfo {author} {\bibfnamefont {J.}~\bibnamefont {Jiang}},
  \bibinfo {author} {\bibfnamefont {T.}~\bibnamefont {Tang}}, \bibinfo {author}
  {\bibfnamefont {M.}~\bibnamefont {Webb}}, \ and\ \bibinfo {author}
  {\bibfnamefont {Y.}~\bibnamefont {Liu}},\ }\bibfield  {title} {\enquote
  {\bibinfo {title} {Dynamics in spinor condensates tuned by a microwave
  dressing field},}\ }\href {\doibase 10.1103/PhysRevA.89.023608} {\bibfield
  {journal} {\bibinfo  {journal} {Phys. Rev. A}\ }\textbf {\bibinfo {volume}
  {89}},\ \bibinfo {pages} {023608} (\bibinfo {year} {2014})}\BibitemShut
  {NoStop}%
\bibitem [{Note1()}]{Note1}%
  \BibitemOpen
  \bibinfo {note} {From the training results of small systems where noises are
  taken into account, we learn that after the system approaches the polar
  state, an additional period of spin mixing dynamics is needed to submerge the
  effect of noise by signal amplification. The duration of this extra evolution
  is optimized in experiments to give the highest metrological
  gain.}\BibitemShut {Stop}%
\bibitem [{\citenamefont {{Schulman}}\ \emph {et~al.}(2017)\citenamefont
  {{Schulman}}, \citenamefont {{Wolski}}, \citenamefont {{Dhariwal}},
  \citenamefont {{Radford}},\ and\ \citenamefont {{Klimov}}}]{ppo2017}%
  \BibitemOpen
  \bibfield  {author} {\bibinfo {author} {\bibfnamefont {John}\ \bibnamefont
  {{Schulman}}}, \bibinfo {author} {\bibfnamefont {Filip}\ \bibnamefont
  {{Wolski}}}, \bibinfo {author} {\bibfnamefont {Prafulla}\ \bibnamefont
  {{Dhariwal}}}, \bibinfo {author} {\bibfnamefont {Alec}\ \bibnamefont
  {{Radford}}}, \ and\ \bibinfo {author} {\bibfnamefont {Oleg}\ \bibnamefont
  {{Klimov}}},\ }\bibfield  {title} {\enquote {\bibinfo {title} {{Proximal
  Policy Optimization Algorithms}},}\ }\href@noop {} {\bibfield  {journal}
  {\bibinfo  {journal} {arXiv e-prints}\ ,\ \bibinfo {eid} {arXiv:1707.06347}}
  (\bibinfo {year} {2017})},\ \Eprint {http://arxiv.org/abs/1707.06347}
  {arXiv:1707.06347 [cs.LG]} \BibitemShut {NoStop}%
\bibitem [{\citenamefont {{He}}\ \emph {et~al.}(2015)\citenamefont {{He}},
  \citenamefont {{Zhang}}, \citenamefont {{Ren}},\ and\ \citenamefont
  {{Sun}}}]{he2015}%
  \BibitemOpen
  \bibfield  {author} {\bibinfo {author} {\bibfnamefont {Kaiming}\ \bibnamefont
  {{He}}}, \bibinfo {author} {\bibfnamefont {Xiangyu}\ \bibnamefont {{Zhang}}},
  \bibinfo {author} {\bibfnamefont {Shaoqing}\ \bibnamefont {{Ren}}}, \ and\
  \bibinfo {author} {\bibfnamefont {Jian}\ \bibnamefont {{Sun}}},\ }\bibfield
  {title} {\enquote {\bibinfo {title} {{Deep Residual Learning for Image
  Recognition}},}\ }\href@noop {} {\bibfield  {journal} {\bibinfo  {journal}
  {arXiv e-prints}\ ,\ \bibinfo {eid} {arXiv:1512.03385}} (\bibinfo {year}
  {2015})},\ \Eprint {http://arxiv.org/abs/1512.03385} {arXiv:1512.03385
  [cs.CV]} \BibitemShut {NoStop}%
\bibitem [{\citenamefont {Achiam}(2018)}]{ppo}%
  \BibitemOpen
  \bibfield  {author} {\bibinfo {author} {\bibfnamefont {Joshua}\ \bibnamefont
  {Achiam}},\ }\bibfield  {title} {\enquote {\bibinfo {title} {Spinning up in
  deep reinforcement learning},}\ }\href@noop {} {\  (\bibinfo {year}
  {2018})}\BibitemShut {NoStop}%
\bibitem [{\citenamefont {Brockman}\ \emph {et~al.}(2016)\citenamefont
  {Brockman}, \citenamefont {Cheung}, \citenamefont {Pettersson}, \citenamefont
  {Schneider}, \citenamefont {Schulman}, \citenamefont {Tang},\ and\
  \citenamefont {Zaremba}}]{gym}%
  \BibitemOpen
  \bibfield  {author} {\bibinfo {author} {\bibfnamefont {Greg}\ \bibnamefont
  {Brockman}}, \bibinfo {author} {\bibfnamefont {Vicki}\ \bibnamefont
  {Cheung}}, \bibinfo {author} {\bibfnamefont {Ludwig}\ \bibnamefont
  {Pettersson}}, \bibinfo {author} {\bibfnamefont {Jonas}\ \bibnamefont
  {Schneider}}, \bibinfo {author} {\bibfnamefont {John}\ \bibnamefont
  {Schulman}}, \bibinfo {author} {\bibfnamefont {Jie}\ \bibnamefont {Tang}}, \
  and\ \bibinfo {author} {\bibfnamefont {Wojciech}\ \bibnamefont {Zaremba}},\
  }\href@noop {} {\enquote {\bibinfo {title} {Openai gym},}\ } (\bibinfo {year}
  {2016}),\ \Eprint {http://arxiv.org/abs/arXiv:1606.01540} {arXiv:1606.01540}
  \BibitemShut {NoStop}%
\bibitem [{\citenamefont {Luo}\ \emph {et~al.}(2017)\citenamefont {Luo},
  \citenamefont {Zou}, \citenamefont {Wu}, \citenamefont {Liu}, \citenamefont
  {Han}, \citenamefont {Tey},\ and\ \citenamefont {You}}]{luo2017}%
  \BibitemOpen
  \bibfield  {author} {\bibinfo {author} {\bibfnamefont {Xin-Yu}\ \bibnamefont
  {Luo}}, \bibinfo {author} {\bibfnamefont {Yi-Quan}\ \bibnamefont {Zou}},
  \bibinfo {author} {\bibfnamefont {Ling-Na}\ \bibnamefont {Wu}}, \bibinfo
  {author} {\bibfnamefont {Qi}~\bibnamefont {Liu}}, \bibinfo {author}
  {\bibfnamefont {Ming-Fei}\ \bibnamefont {Han}}, \bibinfo {author}
  {\bibfnamefont {Meng~Khoon}\ \bibnamefont {Tey}}, \ and\ \bibinfo {author}
  {\bibfnamefont {Li}~\bibnamefont {You}},\ }\bibfield  {title} {\enquote
  {\bibinfo {title} {Deterministic entanglement generation from driving through
  quantum phase transitions},}\ }\href {\doibase 10.1126/science.aag1106}
  {\bibfield  {journal} {\bibinfo  {journal} {Science}\ }\textbf {\bibinfo
  {volume} {355}},\ \bibinfo {pages} {620--623} (\bibinfo {year}
  {2017})}\BibitemShut {NoStop}%
\bibitem [{\citenamefont {Hoang}\ \emph {et~al.}(2016)\citenamefont {Hoang},
  \citenamefont {Bharath}, \citenamefont {Boguslawski}, \citenamefont {Anquez},
  \citenamefont {Robbins},\ and\ \citenamefont {Chapman}}]{hoang2016}%
  \BibitemOpen
  \bibfield  {author} {\bibinfo {author} {\bibfnamefont {Thai~M.}\ \bibnamefont
  {Hoang}}, \bibinfo {author} {\bibfnamefont {Hebbe~M.}\ \bibnamefont
  {Bharath}}, \bibinfo {author} {\bibfnamefont {Matthew~J.}\ \bibnamefont
  {Boguslawski}}, \bibinfo {author} {\bibfnamefont {Martin}\ \bibnamefont
  {Anquez}}, \bibinfo {author} {\bibfnamefont {Bryce~A.}\ \bibnamefont
  {Robbins}}, \ and\ \bibinfo {author} {\bibfnamefont {Michael~S.}\
  \bibnamefont {Chapman}},\ }\bibfield  {title} {\enquote {\bibinfo {title}
  {Adiabatic quenches and characterization of amplitude excitations in a
  continuous quantum phase transition},}\ }\href {\doibase
  10.1073/pnas.1600267113} {\bibfield  {journal} {\bibinfo  {journal} {Proc.
  Natl. Acad. Sci. U.S.A.}\ }\textbf {\bibinfo {volume} {113}},\ \bibinfo
  {pages} {9475--9479} (\bibinfo {year} {2016})},\ \Eprint
  {http://arxiv.org/abs/https://www.pnas.org/content/113/34/9475.full.pdf}
  {https://www.pnas.org/content/113/34/9475.full.pdf} \BibitemShut {NoStop}%
\bibitem [{\citenamefont {Klempt}\ \emph {et~al.}(2010)\citenamefont {Klempt},
  \citenamefont {Topic}, \citenamefont {Gebreyesus}, \citenamefont {Scherer},
  \citenamefont {Henninger}, \citenamefont {Hyllus}, \citenamefont {Ertmer},
  \citenamefont {Santos},\ and\ \citenamefont {Arlt}}]{arlt2010}%
  \BibitemOpen
  \bibfield  {author} {\bibinfo {author} {\bibfnamefont {C.}~\bibnamefont
  {Klempt}}, \bibinfo {author} {\bibfnamefont {O.}~\bibnamefont {Topic}},
  \bibinfo {author} {\bibfnamefont {G.}~\bibnamefont {Gebreyesus}}, \bibinfo
  {author} {\bibfnamefont {M.}~\bibnamefont {Scherer}}, \bibinfo {author}
  {\bibfnamefont {T.}~\bibnamefont {Henninger}}, \bibinfo {author}
  {\bibfnamefont {P.}~\bibnamefont {Hyllus}}, \bibinfo {author} {\bibfnamefont
  {W.}~\bibnamefont {Ertmer}}, \bibinfo {author} {\bibfnamefont
  {L.}~\bibnamefont {Santos}}, \ and\ \bibinfo {author} {\bibfnamefont {J.~J.}\
  \bibnamefont {Arlt}},\ }\bibfield  {title} {\enquote {\bibinfo {title}
  {Parametric amplification of vacuum fluctuations in a spinor condensate},}\
  }\href {\doibase 10.1103/PhysRevLett.104.195303} {\bibfield  {journal}
  {\bibinfo  {journal} {Phys. Rev. Lett.}\ }\textbf {\bibinfo {volume} {104}},\
  \bibinfo {pages} {195303} (\bibinfo {year} {2010})}\BibitemShut {NoStop}%
\bibitem [{\citenamefont {Holland}\ and\ \citenamefont
  {Burnett}(1993)}]{holland1993}%
  \BibitemOpen
  \bibfield  {author} {\bibinfo {author} {\bibfnamefont {M.~J.}\ \bibnamefont
  {Holland}}\ and\ \bibinfo {author} {\bibfnamefont {K.}~\bibnamefont
  {Burnett}},\ }\bibfield  {title} {\enquote {\bibinfo {title} {Interferometric
  detection of optical phase shifts at the heisenberg limit},}\ }\href
  {\doibase 10.1103/PhysRevLett.71.1355} {\bibfield  {journal} {\bibinfo
  {journal} {Phys. Rev. Lett.}\ }\textbf {\bibinfo {volume} {71}},\ \bibinfo
  {pages} {1355--1358} (\bibinfo {year} {1993})}\BibitemShut {NoStop}%
\bibitem [{\citenamefont {Johansson}\ \emph {et~al.}(2013)\citenamefont
  {Johansson}, \citenamefont {Nation},\ and\ \citenamefont
  {Nori}}]{JOHANSSON20131234}%
  \BibitemOpen
  \bibfield  {author} {\bibinfo {author} {\bibfnamefont {J.R.}\ \bibnamefont
  {Johansson}}, \bibinfo {author} {\bibfnamefont {P.D.}\ \bibnamefont
  {Nation}}, \ and\ \bibinfo {author} {\bibfnamefont {Franco}\ \bibnamefont
  {Nori}},\ }\bibfield  {title} {\enquote {\bibinfo {title} {Qutip 2: A python
  framework for the dynamics of open quantum systems},}\ }\href {\doibase
  https://doi.org/10.1016/j.cpc.2012.11.019} {\bibfield  {journal} {\bibinfo
  {journal} {Computer Physics Communications}\ }\textbf {\bibinfo {volume}
  {184}},\ \bibinfo {pages} {1234--1240} (\bibinfo {year} {2013})}\BibitemShut
  {NoStop}%
\end{thebibliography}%
\end{document}